\documentclass[%
 reprint,
superscriptaddress,
 amsmath,amssymb,
 aps,
pra,
]{revtex4-2}

\usepackage{graphicx}
\usepackage{dcolumn}
\usepackage{bm}
\usepackage{hyperref}
\usepackage{physics}
\hypersetup{
  colorlinks=true,
  allcolors=blue,
}

\usepackage{etoolbox,xcolor}

\makeatletter
\pretocmd\frontmatter@thefootnote{\color{blue}}{}{}
\makeatother

\usepackage[caption=false]{subfig}
\newcommand{\phantomsubfloat}[1]{
    {
    \captionsetup[subfloat]{farskip=0pt,captionskip=0pt}
    \captionsetup[subfigure]{labelformat=empty}
    \subfloat{#1}
    }
}

\begin{document}

\title{Optimizing Quantum Algorithms on Bipotent Architectures}

\author{Yanjun Ji}
\email{yanjun.ji@informatik.uni-stuttgart.de}
\affiliation{Institute of Computer Architecture and Computer Engineering, University of Stuttgart, Pfaffenwaldring 47, 70569 Stuttgart, Germany}

\author{Kathrin F. Koenig}
\email{kathrin.koenig@iaf.fraunhofer.de}
\affiliation{Fraunhofer Institute for Applied Solid State Physics IAF, Tullastrasse 72, 79108 Freiburg, Germany}
\affiliation{Faculty of Engineering, University of Freiburg, Georges-Köhler-Allee 101, 79110 Freiburg, Germany}

\author{Ilia Polian}
\email{ilia.polian@informatik.uni-stuttgart.de}
\affiliation{Institute of Computer Architecture and Computer Engineering, University of Stuttgart, Pfaffenwaldring 47, 70569 Stuttgart, Germany}

\date{\today}

\begin{abstract}
Vigorous optimization of quantum gates has led to bipotent quantum architectures, where the optimized gates are available for some qubits but not for others. However, such gate-level improvements limit the application of user-side pulse-level optimizations, which have proven effective for quantum circuits with a high level of regularity, such as the ansatz circuit of the Quantum Approximate Optimization Algorithm (QAOA). In this paper, we investigate the trade-off between hardware-level and algorithm-level improvements on bipotent quantum architectures. Our results for various QAOA instances on two quantum computers offered by IBM indicate that the benefits of pulse-level optimizations currently outweigh the improvements due to vigorously optimized monolithic gates. Furthermore, our data indicate that the fidelity of circuit primitives is not always the best indicator for the overall algorithm performance; also their gate type and schedule duration should be taken into account. This effect is particularly pronounced for QAOA on dense portfolio optimization problems, since their transpilation requires many SWAP gates, for which efficient pulse-level optimization exists. Our findings provide practical guidance on optimal qubit selection on bipotent quantum architectures and suggest the need for improvements of those architectures, ultimately making pulse-level optimization available for all gate types.
\end{abstract}

\maketitle

\section{Introduction}
Computational capabilities of today's quantum architectures are severely limited by comparatively high error rates of their hardware components. This shortcoming is commonly addressed on two levels: by providing improved implementations of qubits and quantum gates, and by adapting a given quantum circuit to error mechanisms and error rates of a specific quantum computer. A recent example from the first category is IBM's direct-CX gate that exhibits a shorter duration and a better fidelity than the previous entangling gate design based on the Echo Cross-Resonance (ECR) principle \cite{jurcevic2021demonstration}. Approaches from the second category include noise-aware transpilation methods \cite{murali2019noise} and pulse-level optimization of algorithm primitives \cite{earnest2021pulse, gokhale2020optimized}.

\begin{figure}[t]
\centering
	\includegraphics[width=\columnwidth]{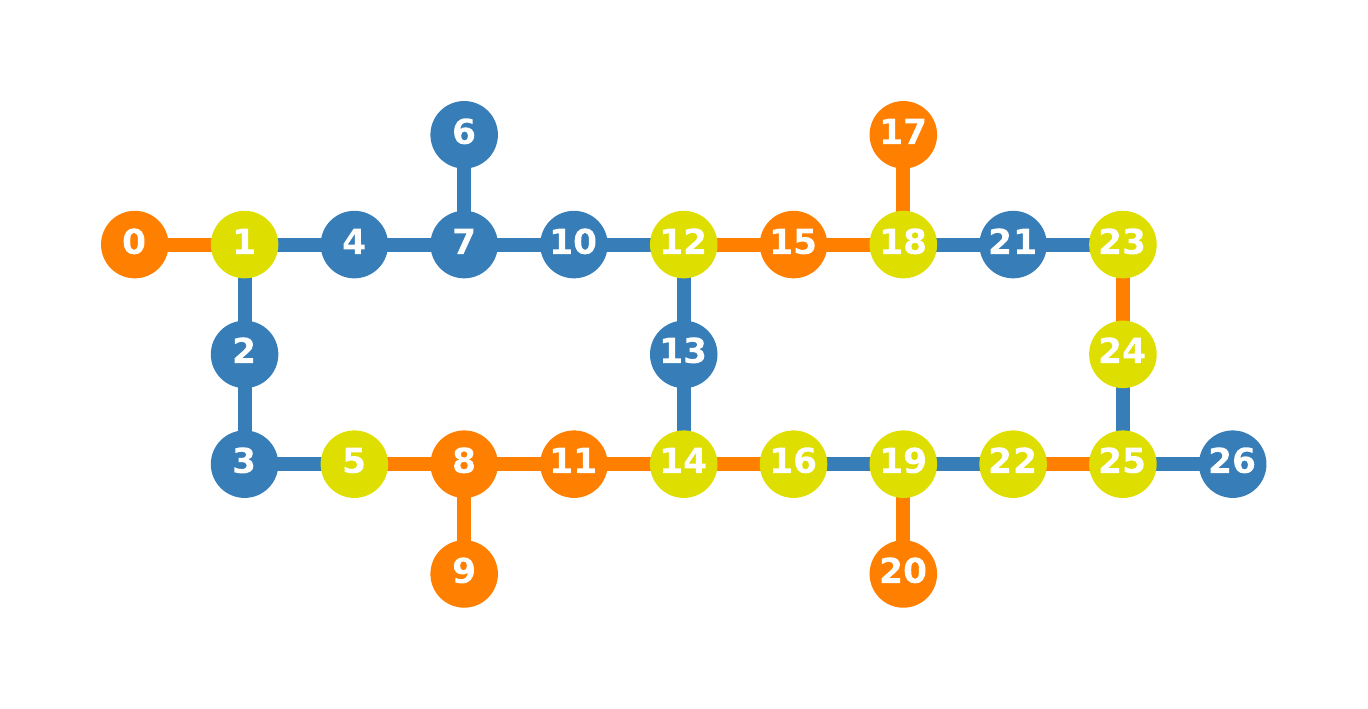}
	\vspace{-20pt}
	\caption{Bipotent architecture of {ibmq\_ehningen}. It contains two types of CX gates: direct-CX, represented by blue or dark gray edges, and ECR-CX, represented by orange or medium gray edges. Qubits are depicted as circles, the lines between the qubits are connections where a CX gate can be applied. When only one type of CX gate can be applied, the qubit has the same color as the CX gate type, otherwise it is green (light gray).}\label{fig:gate_map_ehningen}
\end{figure}

Vigorous optimizations of quantum gates can lead to architectures where the optimized gate is not available for all existing qubits or qubit connections. For example, Fig.~\ref{fig:gate_map_ehningen} shows the topology of the Quantum Processing Unit (QPU) {ibmq\_ehningen}. In this diagram, qubit pairs on which a direct-CX gate implementation is available share a blue (dark gray) edge, whereas qubit pairs on which only the ECR-CX gate implementation is available share an orange (medium gray) edge. While direct-CX gates tend to have better raw error rates, ECR-CX gates offer a significant advantage: they support bespoke pulse-level optimization, which is currently not available for direct-CX gates unless additional calibration is performed. Pulse-level optimizations have proven to be highly effective for regular quantum circuits, such as the Quantum Approximate Optimization Algorithm (QAOA) \cite{farhi2014quantum}, where frequently occurring sub-circuits are mapped to efficient, bespoke sequences of pulses \cite{smith2022summary,earnest2021pulse}.

In this paper, we investigate the qubit mapping of QAOA onto quantum computers that provide two different implementations of the same gate on different pairs of qubits. We call such quantum architectures \emph{bipotent}. Figure~\ref{fig:gate_map_ehningen} is an example of a bipotent architecture, where two different CX gate types are available on different qubit pairs. Similarly, if there are more than two implementations of the same gate, we call such architectures \emph{multipotent}. The reason for adopting a bipotent architecture may be the need to balance the quality of all qubits in engineering or manufacturing and the difficulty of having equally high-quality qubits due to various noises.

The qubit mapping must balance between hardware-level and algorithm-level improvements, that is, trade the lower error rate of direct-CX gates for pulse-level optimization supported by ECR-CX gates. We focus on a specific quantum algorithm: QAOA for Portfolio Optimization (PortOpt) and Maximum-Cut (MaxCut) instances with denser connectivity.

We start by algorithm-agnostically benchmarking direct-CX versus ECR-CX gates, generating data specific to the platforms being used. We then introduce pulse-level optimizations for QAOA circuits applied to PortOpt. In particular, mapping a dense PortOpt problem to a rather sparse topology map such as in Fig.~\ref{fig:gate_map_ehningen} will require many SWAP gates, leading to a substantial contribution to the entire circuit's error rate. Good pulse-level optimizations are known for such SWAP gates. Note that finding optimal pulse sequences is computationally expensive; for example, the compilation time of \emph{GRadient Ascent Pulse Engineering} (GRAPE) scales exponentially in the size of the quantum algorithm \cite{khaneja2005optimal}. For this reason, we only optimize the pulses of frequently occurring primitives of QAOA. The size of the primitives does not scale with the size of the algorithm. Moreover, we focus on improvements that rely on available calibration data and use pulse-level optimizations where possible, but do not consider solutions that require additional calibrations.

We report a comprehensive set of demonstrations on two of IBM's bipotent architectures. For the first time, we investigate not only the fidelities of QAOA circuits, but also their gate types and actual schedule durations on a bipotent architecture. Our results indicate that, for today’s error rates, pulse-level optimization leads to a stronger error-rate decrease than using improved but monolithic native gates in most cases. This comes with a few unexpected observations. For example, the relative impact of pulse-level optimization and monolithic gates varies for circuits of different sizes; pulse-level optimizations tend to outperform direct-CX gates for medium-scale circuits. Moreover, the Control-Target polarity of CX gates, which is usually ignored by implementing algorithms, has a significant influence on the quality of the results.

The key contributions and novelties of this paper are as follows.
\begin{itemize}
    \item For the first time, the features of a bipotent architecture are studied and used to optimize the implementation of algorithms on the QPUs.
    \item We improve the performance of QAOA for the first time by selectively optimizing partial gates implemented on some specific instead of all qubit pairs. This optimization technique can also be applied to other quantum algorithms.
    \item We describe how vigorously optimized quantum gates lead to bipotent quantum architectures and identify the conflict between using highly calibrated monolithic gates on the one hand, and pulse-level optimization on the other.
    \item We extensively study the performance of QAOA on noisy quantum computers for different problem sizes and depths using different strategies for selecting direct-CX or/and ECR-CX gates.
    \item We provide practical advice on how to map QAOA instances onto a given bipotent quantum architecture, thus improving the performance of QAOA on said architectures.
    \item All methods presented in this paper do not require any additional calibration.
\end{itemize}

The remainder of this article is organized as follows. Section \ref{sec:background} reviews the implementation of QAOA on Noisy Intermediate-Scale Quantum (NISQ) devices and the challenges of noise. Section \ref{sec:dcx-ecrcx} provides the features of a bipotent architecture. Section \ref{sec:pulse-lev} discusses pulse-level optimizations in QAOA. Section \ref{sec:bench} demonstrates the performance of QAOA with different optimizations on two QPUs. In Sec. \ref{sec:discu}, we discuss the case beyond QAOA and error mitigation techniques. Finally, Sec. \ref{sec:concl} concludes this paper. Additional details about the QPUs used in this study are presented in the Appendix.

\section{Background}
\label{sec:background}

\subsection{NISQ computing}
Quantum computers form a new paradigm of computing that is directly based on the laws of quantum mechanics \cite{schwabl2007quantum}. The basic unit of a quantum computer, the qubit, differs from the classical bit because it can be in a quantum \emph{superposition} of the states 0 and 1, and can furthermore be \emph{entangled} with other qubits. These quantum mechanical effects make quantum computers more powerful than classical computers in theory. There are various physical realizations of quantum computers already existing today, such as those based on superconducting circuits \cite{clarke2008superconducting}, photonics \cite{o2009photonic}, quantum dots \cite{jacak2013quantum}, trapped ions \cite{bruzewicz2019trapped} and neutral atoms \cite{henriet2020quantum}.

However, it shows to be experimentally extremely challenging to shield qubits from noise. Furthermore, technology has not advanced to the point that the errors induced by this noise can be fully detected and corrected during a quantum computation. Additionally, the currently available quantum computers have a limited number of qubits, and many platforms, including superconducting transmons considered in this work, offer restricted connectivity, i.e., for every qubit, two-qubit gates are only possible with a limited number (typically 2 to 4) of other qubits. These limited devices are commonly referred to as NISQ devices \cite{preskill2018quantum}.

\begin{figure*}[ht]
	\raggedright
	\includegraphics[width=0.14\linewidth]{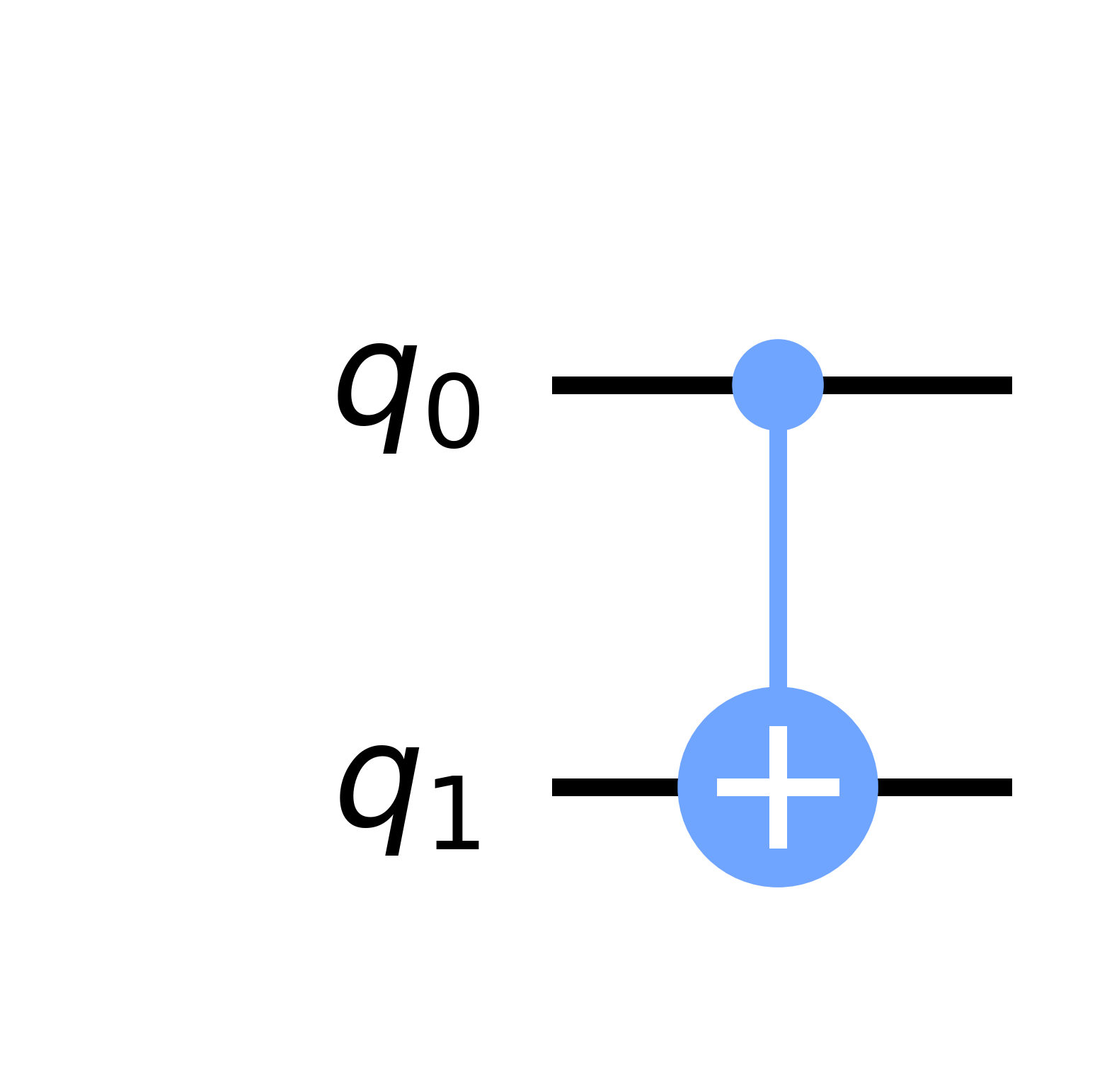}\hfill
	\put(-68,48){\textbf{(a)}}
	\phantomsubfloat{\label{fig:two_cx_gates_a}}
	\includegraphics[width=0.18\linewidth]{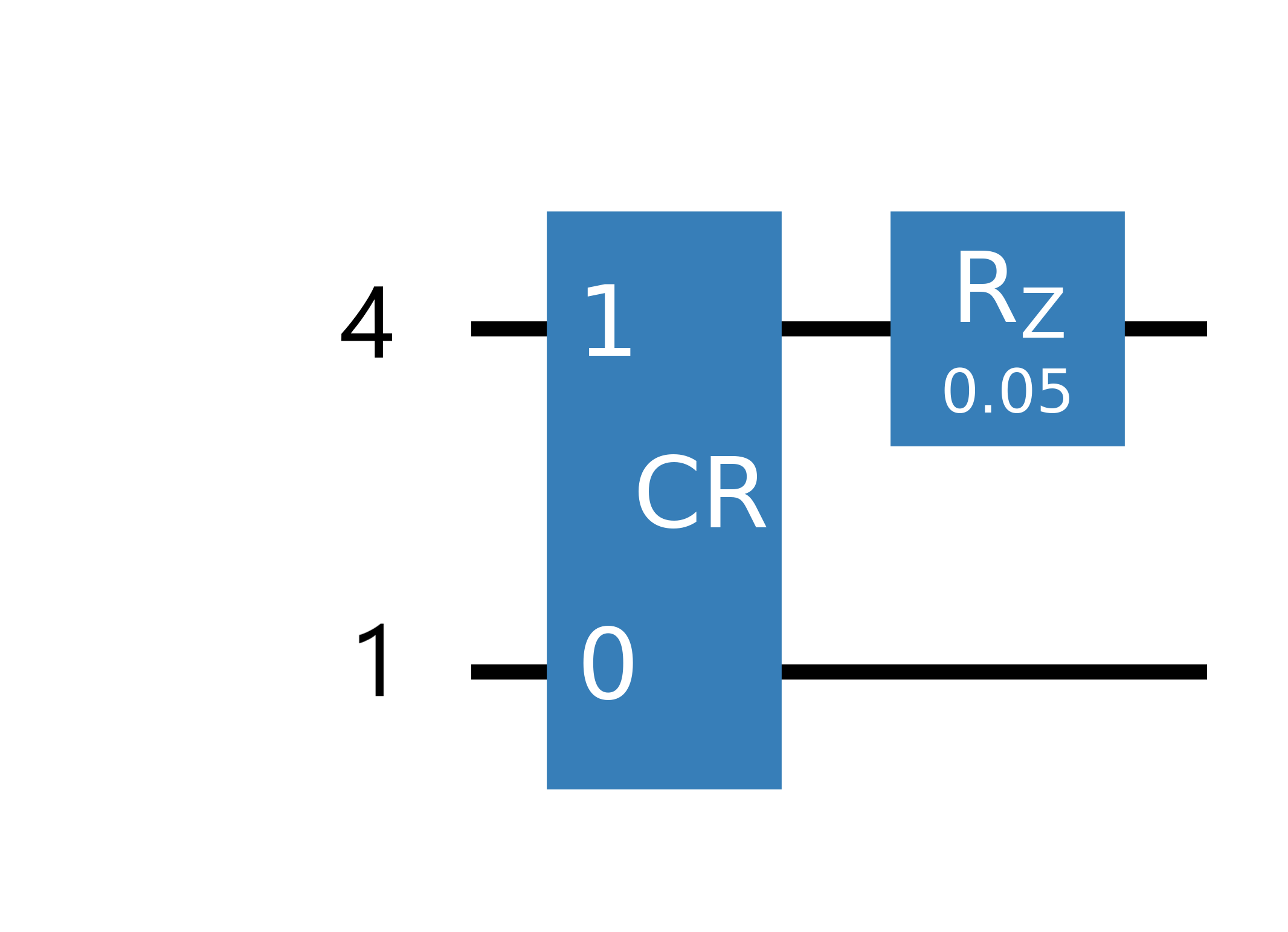}\hfill
	\put(-88,48){\textbf{(b)}}
	\phantomsubfloat{\label{fig:two_cx_gates_b}}
	\includegraphics[width=0.22\linewidth]{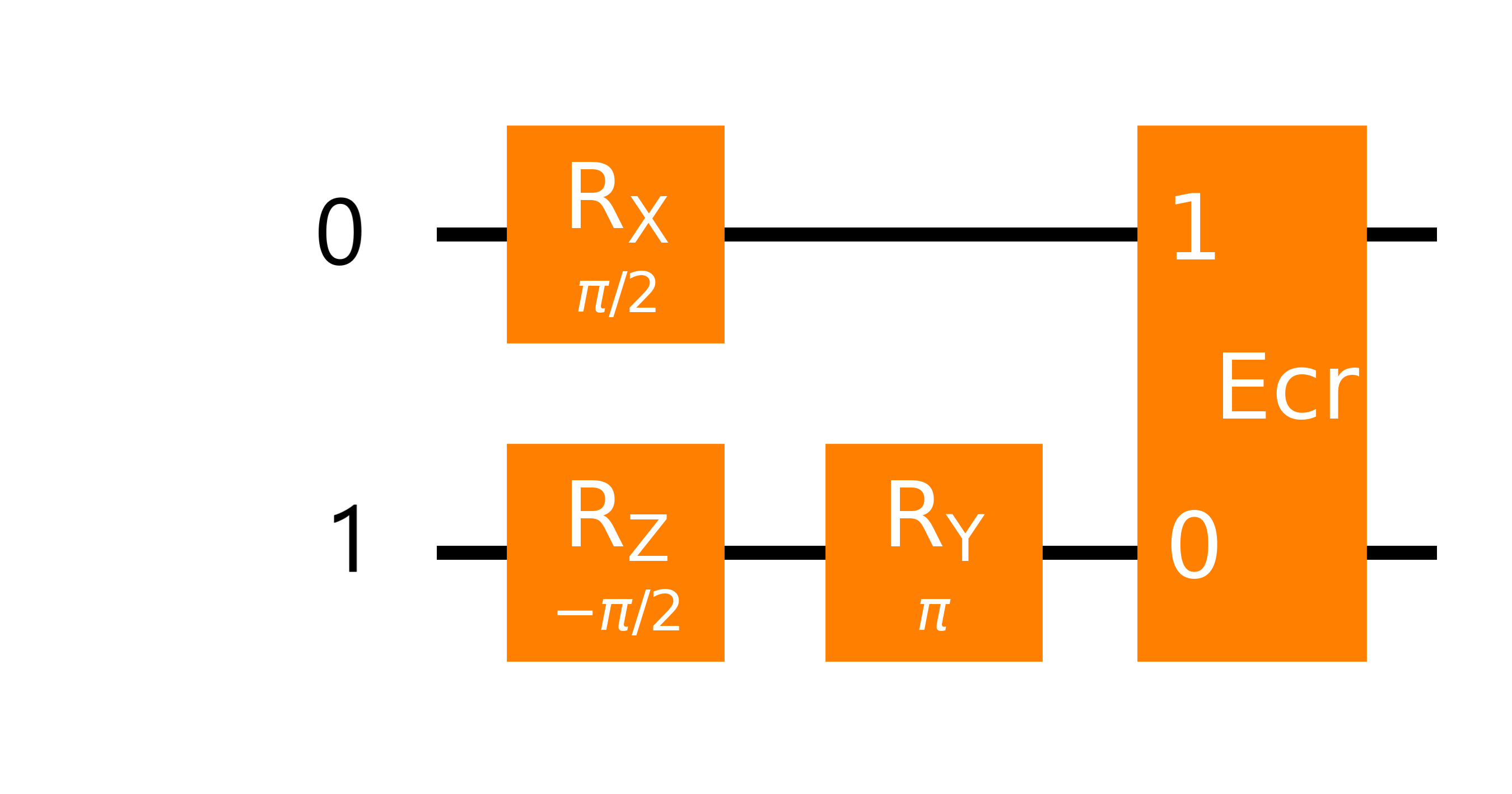}\hfill
	\put(-108,48){\textbf{(c)}}
	\phantomsubfloat{\label{fig:two_cx_gates_c}}
	\includegraphics[width=0.40\linewidth]{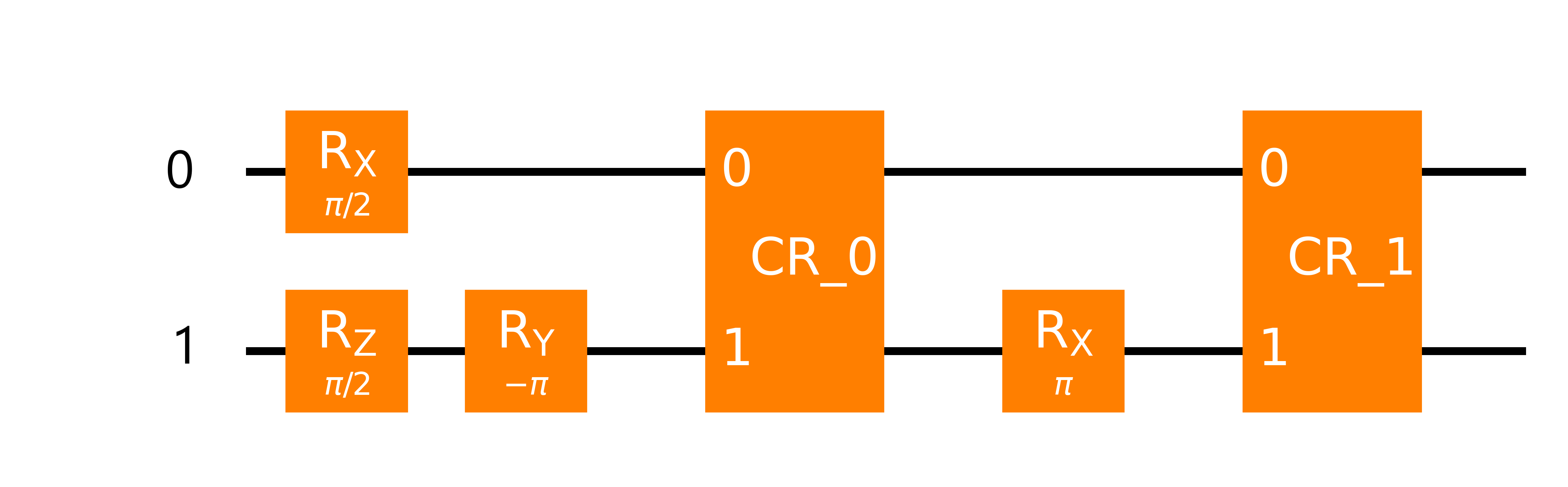}\hfill
	\put(-201,48){\textbf{(d)}}
	\phantomsubfloat{\label{fig:two_cx_gates_d}}
	\vspace{-20pt}
	
	\includegraphics[width=0.23\linewidth]{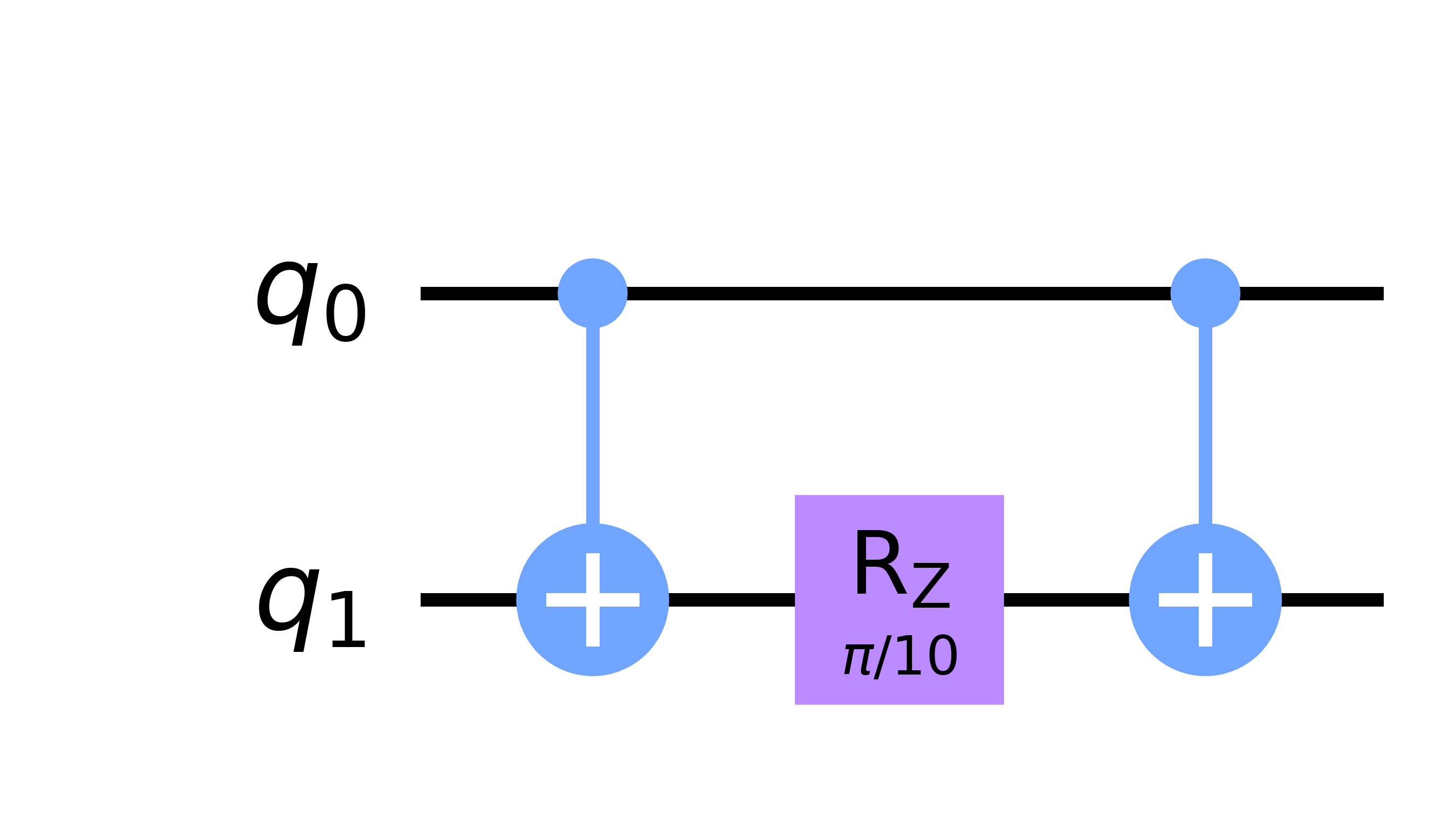}\hfill
	\put(-113,48){\textbf{(e)}}
	\phantomsubfloat{\label{fig:zz_gates_a}}
	\includegraphics[width=0.75\linewidth]{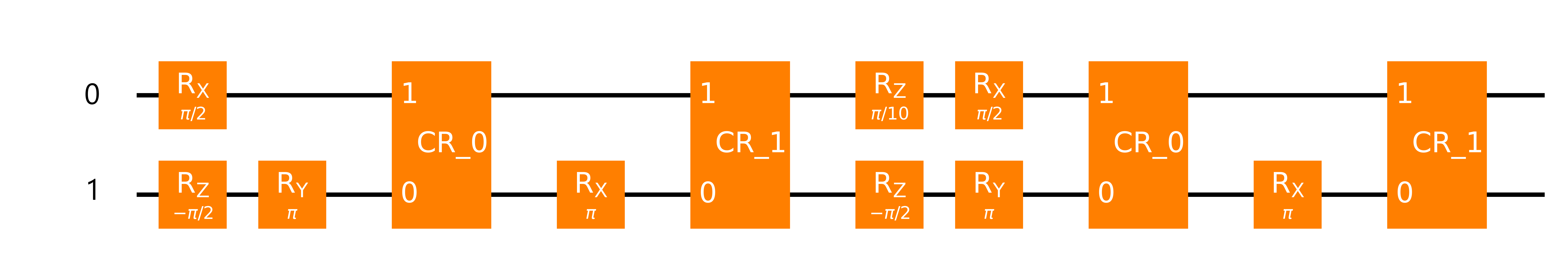}\hfill
	\put(-380,48){\textbf{(f)}}
	\phantomsubfloat{\label{fig:zz_gates_c}}
	\vspace{-20pt}
	
	\phantomsubfloat{\label{fig:zz_gates_b}}
	\includegraphics[width=0.29\linewidth]{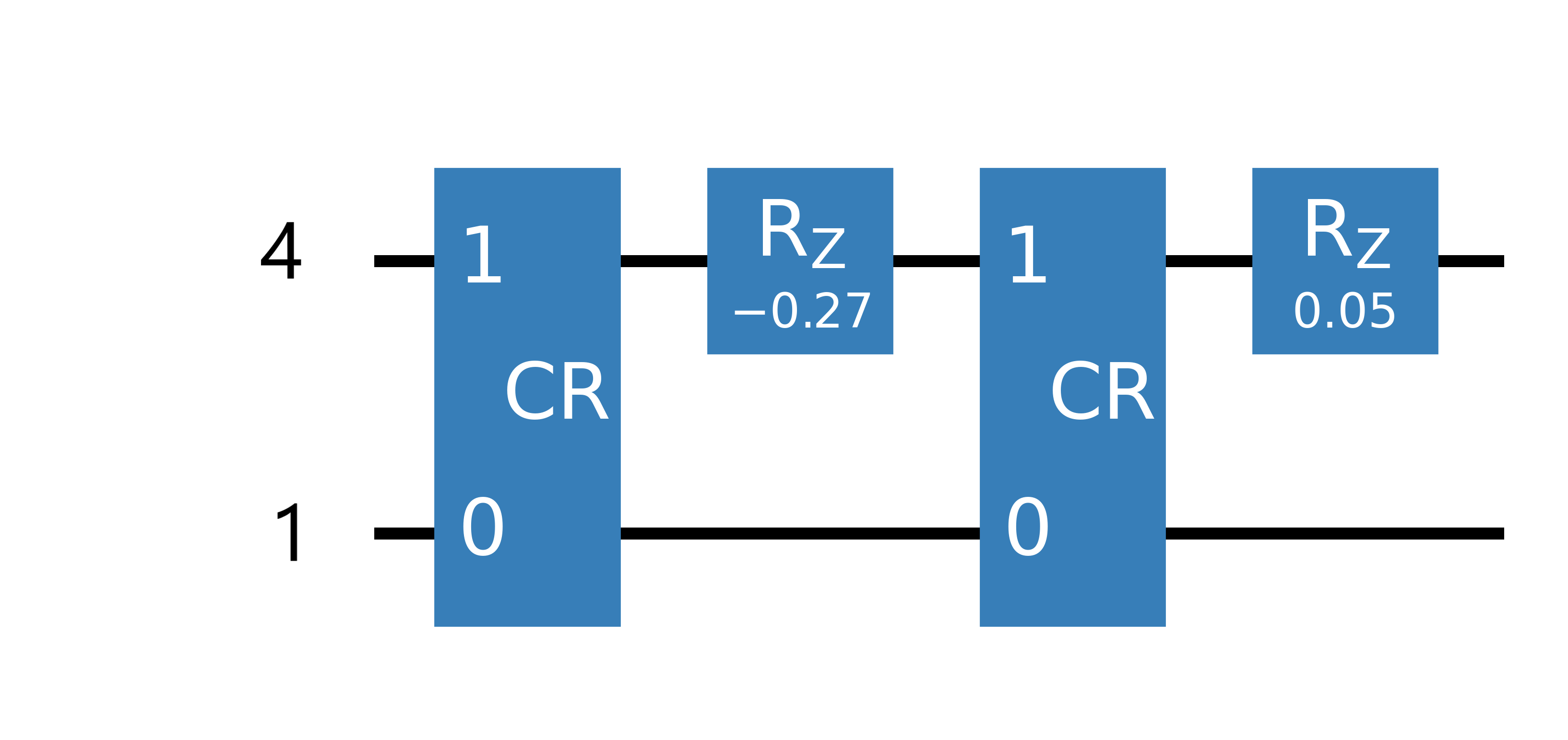}\hfill
	\put(-147,48){\textbf{(g)}}
	\phantomsubfloat{\label{fig:zz_gates_d}}
	\includegraphics[width=0.68\linewidth]{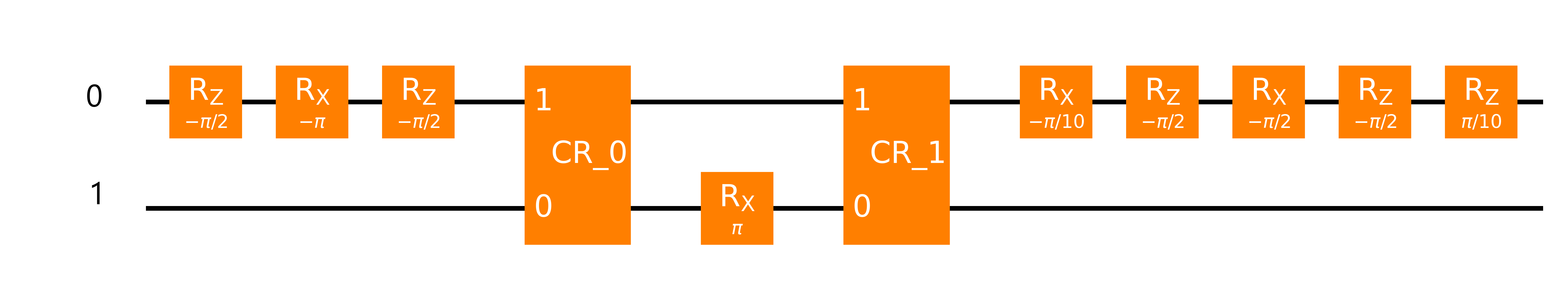}\hfill
	\put(-350,48){\textbf{(h)}}
	\caption{Different implementations of CX and ZZ gates on IBM's QPUs. (a) Symbol of CX gate. (b) direct-CX gate on qubits [1, 4]. (c) ECR-CX gate on qubits [1, 0]. (d) Hardware implementation of the ECR-CX gate. (e) Decomposition of ZZ gate. ZZ gate implemented by (g) direct-CX gates on qubits [1, 4] and (f) ECR-CX gates on qubits [1, 0]. (h) ZZ\textsubscript{OPT} gate, the pulse-level optimization of (f).}
\end{figure*}

Despite their limitations, NISQ devices may already solve problems that are intractable for any classical computers, and hence provide a so-called \emph{quantum advantage}. Because of the noise, it is paramount to keep the quantum circuit depth of any NISQ algorithm to an absolute minimum, which becomes challenging because of the limited connectivity. Furthermore, the average number of errors incurred by the noisy two-qubit gates differs starkly from qubit pair to qubit pair. Hence, in the compilation of quantum algorithms, there should be a preference for executing gates on the pairs that induce few errors.

\subsection{Noise-aware transpilation}

In general, quantum circuits might include gates chosen by the designer and involving arbitrary qubits, whereas actual QPUs offer only a restricted set of quantum gates, called \emph{basis gates}, and limited qubit connectivity. Therefore, a circuit must be \emph{transpiled} to a given QPU: its operation must be mapped to basis gates and connectivity violations must be addressed by, e.g., adding SWAP gates. Different qubits of a QPU are affected by noise to a different extent, and transpilation should be \emph{noise-aware}, i.e., evaluate current calibration data and select the qubits with the highest quality. It is worth noting that, for the considered QPUs, the added SWAP gates contribute to noise significantly, because each such gate includes three CX gates, which are a dominant error source. Therefore, transpilation must balance between selecting better qubits and requiring fewer additional SWAP gates.

Mapping the algorithm's qubits to quantum computer's qubits (the qubit mapping problem) has been shown to be NP-hard. Various transpilation approaches \cite{ji2022calibration, Sivarajah2020tketAR,amy2020staq,harrigan2021quantum,steiger2018projectq,murali2019full,KhammassiASNKRL22,PennyLane} were designed. Furthermore, many heuristic \cite{siraichi2019qubit, zulehner2018efficient, murali2019noise, tannu2019not,  childs19, li2020qubits, niu2020hardware} and exact \cite{siraichi2019qubit,BhattacharjeeC17,ShafaeiSP14,Pedram16,murali2019noise,TanC20, ji2022algorithm} methods were developed to solve qubit mapping problem. While heuristics compile the circuit faster, the exact methods produce solutions that are optimal with respect to a user-specified objective (e.g., circuit depth or fidelity).

In this paper, we use an exact and scalable approach \cite{ji2022algorithm} on a linear topology aiming to minimize the circuit depth to pre-transpile the circuit to satisfy the connectivity constraints. Then, we map the pre-transpiled circuit to different qubits according to the circuit fidelity, gate type, and schedule duration.

\subsection{Quantum gate design and pulse-level optimization}

Quantum gates are a crucial component of quantum algorithms, and their quality is paramount to algorithm performance. Researchers have explored various methods for enhancing gate quality, including investigating different measurement durations to improve readout fidelity \cite{ji2022modellierung} and implementing pulse-level optimizations to enhance gate fidelity. For example, GRAPE can be used to translate quantum algorithms directly into optimal hardware control pulses with shorter pulse lengths than a gate-based compilation, thereby improving fidelity, but with exponentially increasing compilation times. A partial compilation of variational algorithms \cite{gokhale2019partial} was developed to achieve the pulse speedups of GRAPE by precomputing optimal pulses for parametrization-independent blocks of gates. In addition, aggregating small gates into larger operations can reduce the compilation latency \cite{shi2019optimized}.

Compared to directly optimizing algorithms that require extensive computation, optimizing specific gates can improve the performance of the algorithm more effectively. In Ref. \cite{earnest2021pulse}, the authors show a pulse-efficient transpilation approach that requires no additional calibration efforts by scaling Cross-Resonance (CR) pulses and removing redundant single-qubit operations to reduce the overall schedule duration of the gate and thus improves its fidelities. Faster and more reliable SWAP gates were developed at pulse-level with CR native gates on IBM's QPUs \cite{gokhale2021faster}. Moreover, using hardware primitive gates can reduce errors and run times resulting in improved performance \cite{gokhale2020optimized}. In addition, a hybrid gate-pulse model \cite{liang2022hybrid} was proposed to improve the performance of QAOA on IBM's QPUs. Recently, the authors in Ref.~\cite{egger2023study} demonstrated an improvement in performance by achieving shorter pulse schedules through the optimization of pulse amplitude and duration.

\begin{figure}[t]
	\includegraphics[width=\linewidth]{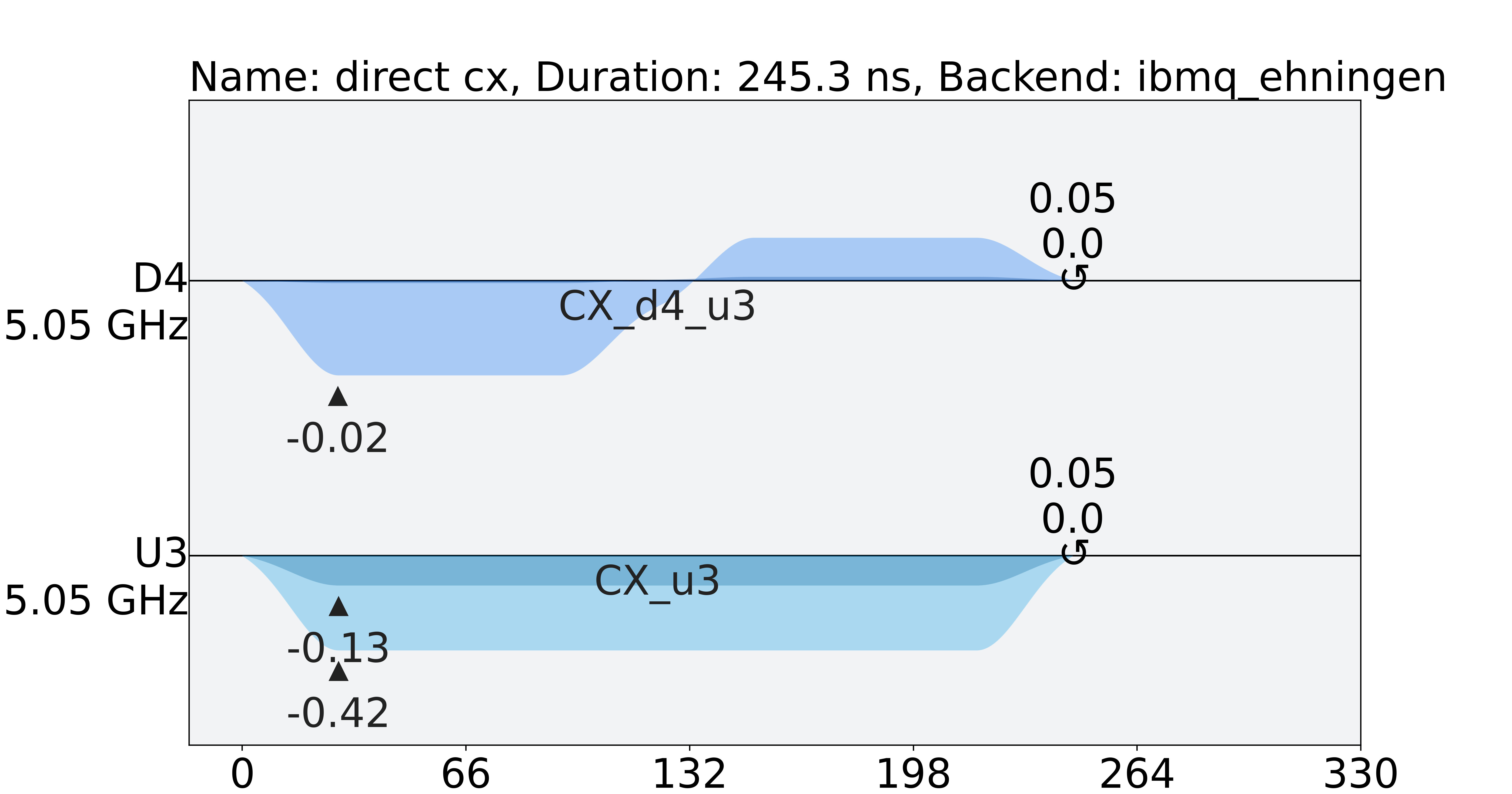}\hfill
	\put(-248,115){\textbf{(a)}}
	\phantomsubfloat{\label{fig:cx_scheds_a}}
	\vspace{-10pt}
	
	\phantomsubfloat{\label{fig:cx_scheds_b}}
	\includegraphics[width=\linewidth]{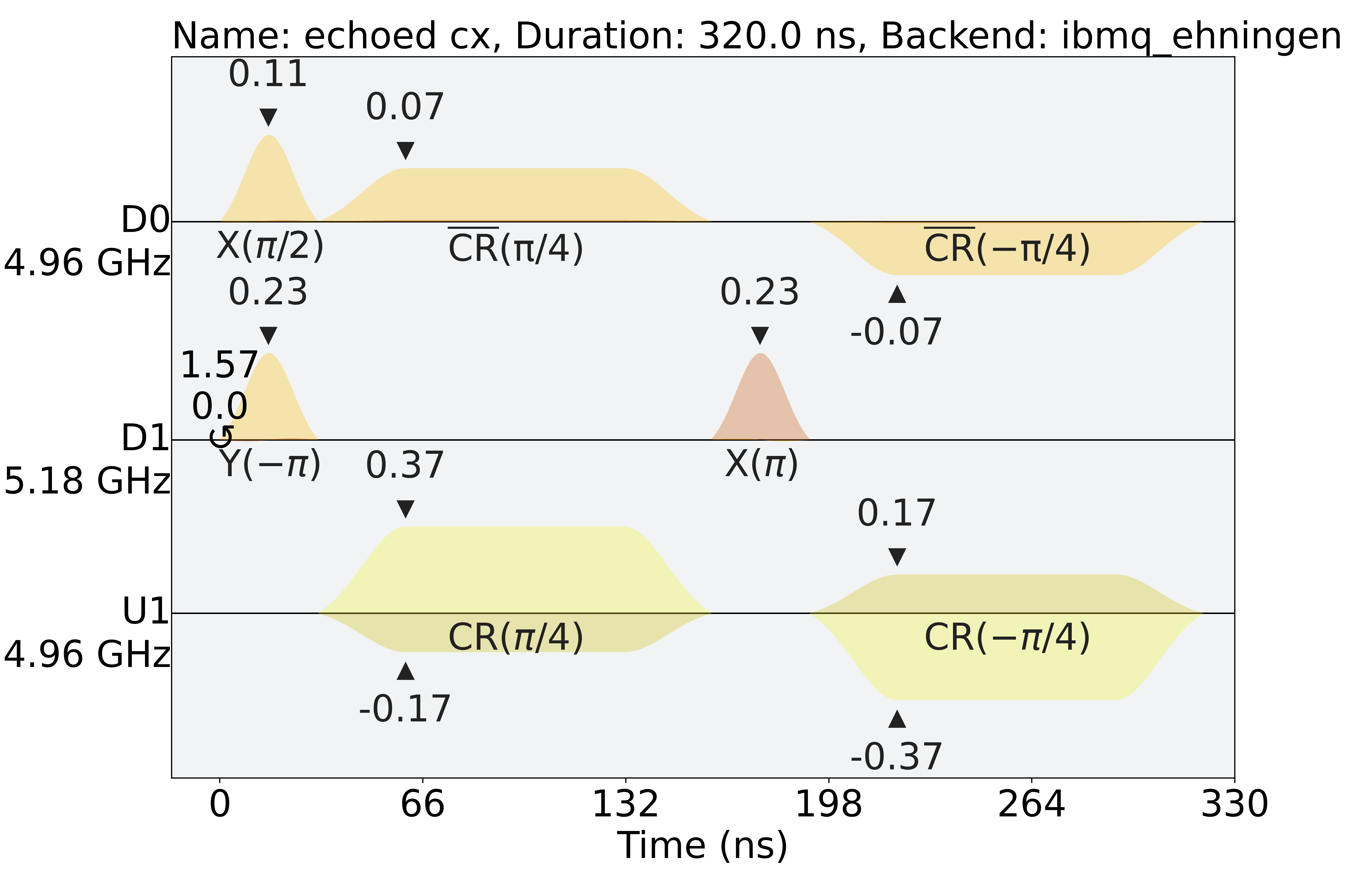}\hfill
	\put(-248,150){\textbf{(b)}}
	
	\caption{Schedules of (a) direct-CX on qubits [1, 4] and (b) ECR-CX on qubits [1, 0].}\label{fig:cx_scheds}
\end{figure}

We now introduce the pulse-level structure of CX gates. Figure~\ref{fig:two_cx_gates_a} is a general symbol of the CX gate on IBM's QPUs. The direct-CX and ECR-CX gates are indistinguishable at the gate-level. However, there are differences in the pulse-level. Figures~\ref{fig:two_cx_gates_b} and \ref{fig:two_cx_gates_c} show the implementations (up to a global phase) of a direct-CX gate on qubits [1, 4] and an ECR-CX gate on qubits [1, 0], respectively. While the direct-CX consists of one CR gate and one RZ gate, the ECR-CX consists of one ECR gate and three single-qubit gates. A closer look into the hardware implementation, the ECR gate is implemented by two CR gates and one RX gate on the control qubit, as shown in Fig.~\ref{fig:two_cx_gates_d}. Their corresponding schedules consisting of several microwave pulses are shown in Fig.~\ref{fig:cx_scheds}. The direct-CX with a schedule duration of $245.3$\,ns is implemented by one Gaussian square pulse on the connecting channel U3, a parallel echoed Gaussian square pulse on the drive channel D4 of target qubit 4, and a virtual (digital) RZ rotation (denoted by a $\circlearrowleft$). In comparison, the ECR-CX gate with a duration of $320$\,ns consists of two CR pulses and four single-qubit gates. The schedule duration of the virtual RZ gate is $0$\,ns, while that of RY and RX gates is $32$\,ns.

Although direct-CX has a shorter schedule duration than ECR-CX, gates implemented by ECR-CX gates may perform better than those implemented by direct-CX, since they can be further optimized directly at the pulse-level by employing existing techniques in Qiskit \cite{qiskit}. Specifically, the ZZ gate implemented by ECR-CX gates allows pulse-level optimization without any additional calibration \cite{earnest2021pulse}. Compared to this, optimizing gates implemented by direct-CX gates currently requires additional calibration (e.g., \cite{Peterson2022optimalsynthesis}). Figures~\ref{fig:zz_gates_a}-\ref{fig:zz_gates_d} show the different implementations of ZZ gate on IBM's QPUs. It achieves a schedule duration of $490$\,ns on qubits [1, 4] using direct-CX gates, compared to a value of $640$\,ns on qubits [1, 0] using ECR-CX gates. However, the ZZ gate implemented by ECR-CX gates can be optimized at pulse-level resulting in a reduced duration of $241.8$\,ns. We refer to the ZZ gate optimized at pulse-level as ZZ\textsubscript{OPT}. In addition, single-pulse-gates RX and RY in ECR-CX enable gate cancellations, while a virtual RZ gate with 0\,ns schedule duration in direct-CX can not be further optimized.

As shown in Ref. \cite{jurcevic2021demonstration}, the direct-CX gate seems to have better quality. At the same time, the ECR-CX gate gives the user flexibility for pulse-level optimization \cite{earnest2021pulse}, which will be discussed in Sec. \ref{sec:dcx-ecrcx}.

\subsection{QAOA}
\label{sec:background-qaoa}

Variational Quantum Algorithms (VQAs) hold promise for quantum advantage and have been developed in many applications \cite{cerezo2021variational}. The QAOA is among the most important VQAs and is promising for solving combinatorial optimization problems on NISQ devices \cite{farhi2014quantum}. In this section, we describe the implementation of QAOA for Portfolio Optimization (PortOpt) and MaxCut problems.

\subsubsection{QAOA for PortOpt}

The PortOpt aims to select the best portfolio from all portfolios in order to maximize expected returns and minimize financial risk. The QAOA holds promise for solving this problem \cite{baker2022wasserstein, brandhofer2023benchmarking, egger2020quantum, hegade2022portfolio}.

The PortOpt problem needs to be first transformed into a \emph{Quadratic Unconstrained Binary Optimization} (QUBO) \cite{glover2022quantum}, an NP-hard problem. The PortOpt with the QUBO representation is then converted to quantum operators so that it can be executed by NISQ computers.

Given $n$ available assets and $B$ assets to be selected, the cost function $C$ can be represented in terms of quantum operators \cite{brandhofer2023benchmarking}:
\begin{equation}
\hat{C}= \sum_{i=1}^{n-1}\sum_{j=i+1}^n \frac{\lambda}{2}(q\sigma_{ij}+A)\hat{Z}_i\hat{Z}_j-\sum_{i=1}^n k_i \hat{Z}_i,
\label{eq:cost-function-operator}
\end{equation}
with
\begin{equation*}
k_i=\frac{\lambda}{2} \left[A (2B-n)+(1-q)\mu_i-q \sum_{j=1}^n \sigma_{ij}\right],
\end{equation*}
where $\lambda$, $q$, $\sigma_{ij}$, $A$, and $\mu_i$ are the global scaling factor, risk preference, covariance matrix, penalty factor, and expected return, respectively. $\hat{Z}_i \hat{Z}_j$ and $\hat{Z}_i$ represent the $ZZ$ interaction on qubits [$i$, $j$] and Pauli $Z$ operator acting on qubit $i$, respectively.

The implementation of QAOA starts from an initial state generated by Hadamard gates. After that, ZZ gates are performed with the definition:
\begin{equation}
    ZZ(\gamma) = e^{-i \frac{\gamma}{2} Z\otimes Z},
\end{equation}
where $\gamma$ is the rotation angle, followed by single-qubit rotation gates RZ and RX. To optimize the parameters in QAOA, we utilize the COBYLA optimizer and Qiskit's QASM-Simulator.

To evaluate the performance of QAOA for PortOpt, we define the Approximation Ratio (AR) and Success Probability (SP). A higher value of AR or SP implies better performance. The post-selected results that satisfy the budget constraint (all feasible solutions) are used to calculate AR of $n$ assets $\{z_1, ..., z_n\}$, defined as follows:
\begin{equation}
r(z_1,\dots,z_n)=C(z_1,\dots,z_n)/C_{\rm opt},
\end{equation}
where $C(z_1, ..., z_n)$ and $C_{\rm opt}$ are the mean value found by QAOA and optimal value, respectively. The SP is defined as the probability of optimal solution.

\subsubsection{QAOA for MaxCut}

Another application of QAOA is to find the maximum cut of a graph, i.e., to partition the nodes of the graph into two sets such that the number of edges  between the sets is maximized, which is known as the MaxCut problem \cite{farhi2014quantum, wurtz2021maxcut, chandarana2022digitized}.

The objective function $C$ can be reformulated as quantum operators:
\begin{equation}
    \hat{C} = \frac{1}{2} \sum_{i,j} (1-\hat{Z}_i \hat{Z}_j),
    \label{eq:cf_maxcut}
\end{equation}
where vertices $i$, $j$ share an edge. For a given graph with $n$ edges $\{z_1, ..., z_n\}$, the AR of QAOA for MaxCut is defined by:
\begin{equation}
    r(z_1, ..., z_n) = C(z_1, ..., z_n)/C_{{\rm max}},
\end{equation}
with $C(z_1, ..., z_n)$ the mean value found by QAOA and $C_{\text{max}} = \max C(z_1, ..., z_n)$ the optimal value.

\section{direct-CX vs.~ECR-CX gates}
\label{sec:dcx-ecrcx}

We now investigate the features of a bipotent architecture, namely {ibmq\_ehningen}, as shown in Fig.~\ref{fig:gate_map_ehningen}, on which two types of CX gates are present: direct-CX and ECR-CX. Additionally, we denote the qubits as Q-ECR (orange, medium gray), Q-Direct (blue, dark gray), and Q-Bipotent (green, light gray), depending on whether the qubit is connected to only ECR-CX, only direct-CX, or both.

\begin{figure*}[t]
	\includegraphics[width=\linewidth]{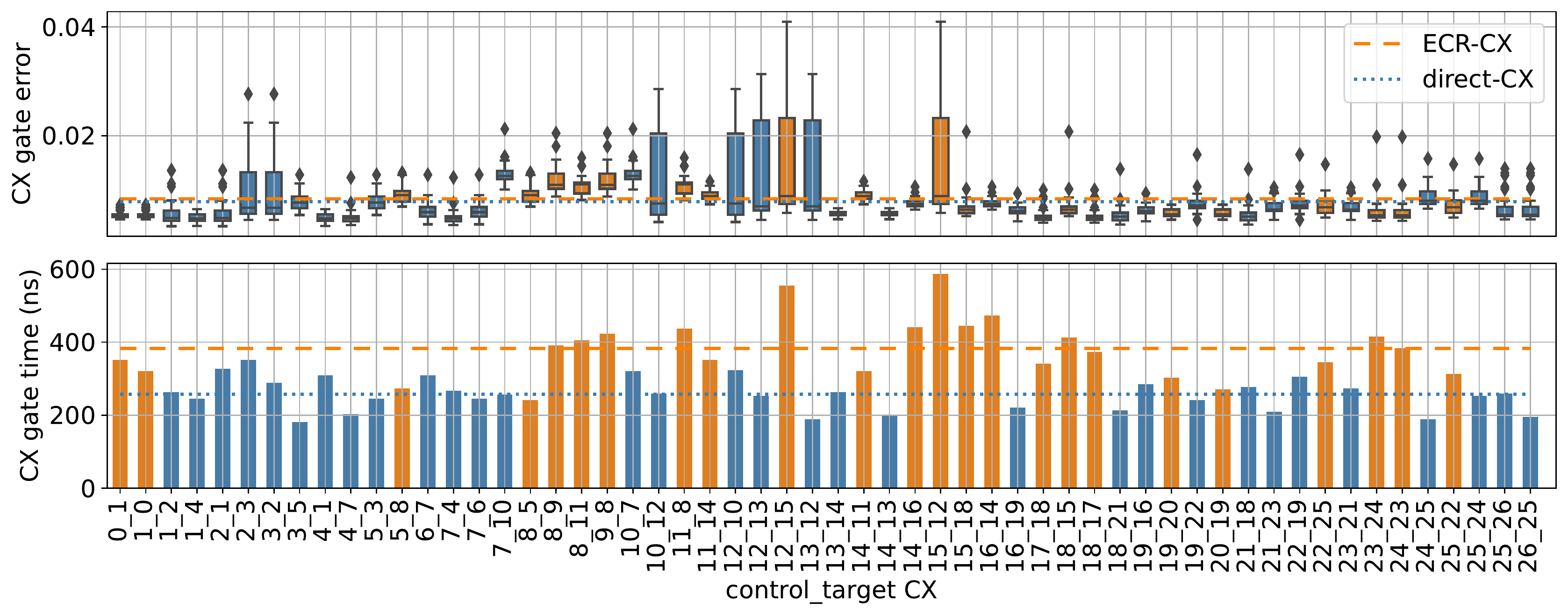}\hfill
	\put(-473,186){\textbf{(a)}}
	\put(-473,105){\textbf{(b)}}
	\phantomsubfloat{\label{fig:cx_all_ehningen_a}}
	\phantomsubfloat{\label{fig:cx_all_ehningen_b}}
	\vspace{-5pt}
	\caption{Properties of ECR-CX (orange, medium gray) and direct-CX (blue, dark gray) gates on {ibmq\_ehningen}. (a) Gate error rate. (b) Execution time in nanoseconds.}\label{fig:cx_all_ehningen}
\end{figure*}

\begin{figure*}[t]
	\phantomsubfloat{\label{fig:t1t2_ehningen_a}}
	\phantomsubfloat{\label{fig:t1t2_ehningen_b}}
	\includegraphics[width=\linewidth]{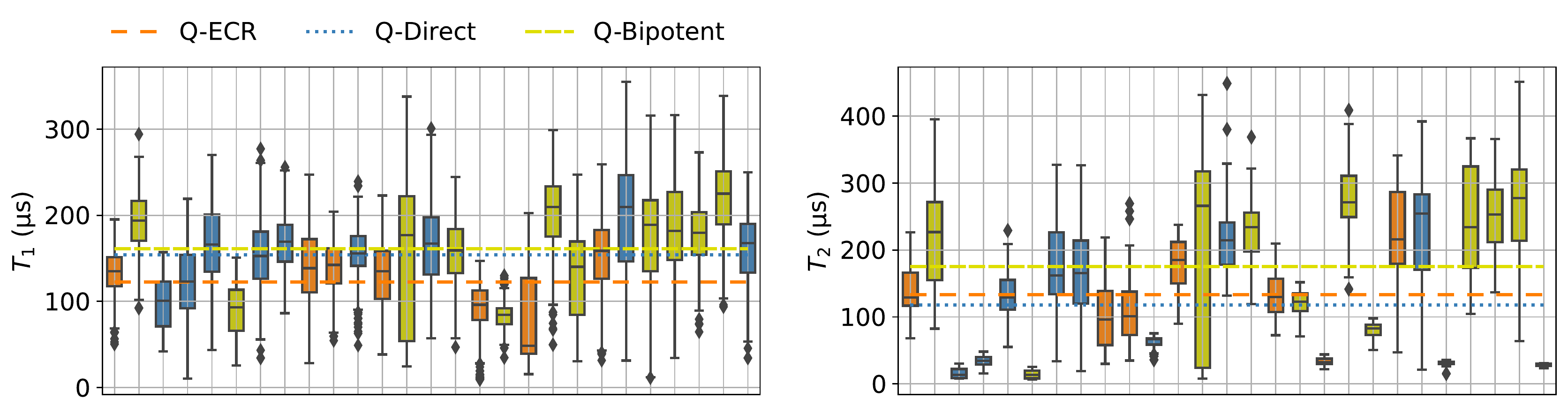}\hfill
	\put(-475,102){\textbf{(a)}}
	\put(-215,102){\textbf{(b)}}
	\vspace{-15pt}
	
	\phantomsubfloat{\label{fig:single_error_all_a}}
	\phantomsubfloat{\label{fig:single_error_all_b}}
	\includegraphics[width=\linewidth]{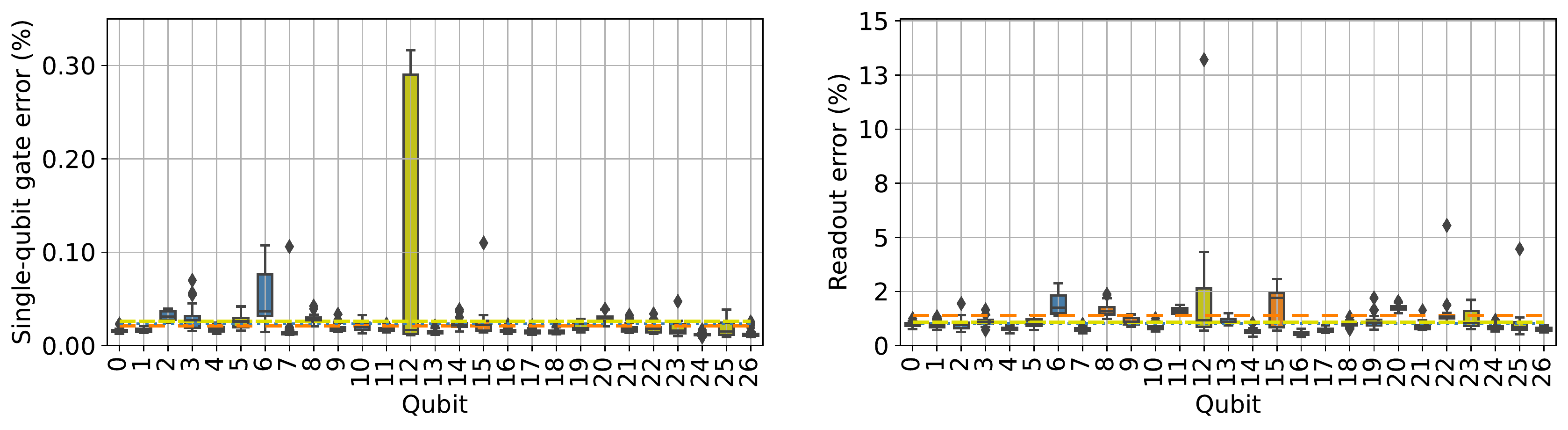}\hfill
	\put(-472,123){\textbf{(c)}}
	\put(-215,123){\textbf{(d)}}
	\caption{Properties of 27 qubits on {ibmq\_ehningen} used for experiments in this article. (a) $T_1$ and (b) $T_2$ decoherence times. A higher value is better. (c) Single-qubit gate error and (d) readout error. A lower value is better.  We label qubits connected to two types of CX gates as Q-Bipotent (green, light gray) and qubits connected only to ECR-CX or direct-CX gates as Q-ECR (orange, medium gray) or Q-Direct (blue, dark gray), respectively.}\label{fig:single_qubit_prop}
\end{figure*}

We collected hourly calibration data of {ibmq\_ehningen} throughout September 2022. Figure~\ref{fig:cx_all_ehningen} shows the CX gate error rate and gate execution time of ECR-CX and direct-CX. While the error rate varies over time, the gate time is constant. Their average values are summarized in Table~\ref{table:two-qubit}. Compared to ECR-CX, the average error rate of direct-CX is slightly reduced by $4.82\%$, while the average gate time is significantly reduced by $32.79\%$. The data show that direct-CX is generally of better quality than ECR-CX.
\begin{table}[b]
		\begin{ruledtabular}
			\centering
			\caption{Properties of ECR-CX and direct-CX gates.}
			\label{table:two-qubit}
			\begin{tabular}{llll}
				\textbf{Parameter} & \textbf{ECR-CX} & \textbf{direct-CX} & \textbf{Reduction (\%)}\\
				Gate error (\%) & $0.83$ & $0.79$ & $4.82$\\
				Gate time (ns) & $382.22$ & $256.89$ & $32.79$\\
			\end{tabular}
		\end{ruledtabular}
		\begin{ruledtabular}
			\centering
			\caption{Properties of qubits.}
			\label{table:single-qubit}
			\begin{tabular}{llll}
				\textbf{Parameter} & \textbf{Q-ECR} & \textbf{Q-Direct} & \textbf{Q-Bipotent}\\
				$T_1~(\rm{\mu s})$ & $122.59$ & $154.29$ & $161.16$ \\
				$T_2~(\rm{\mu s})$ & $132.94$ & $118.02$ & $175.16$ \\
				Gate error (\%) & $0.021$ & $0.023$ & $0.026$ \\
				Readout error (\%) & $1.386$ & $1.009$ & $1.077$ \\
			\end{tabular}
		\end{ruledtabular}
\end{table}

Figures~\ref{fig:t1t2_ehningen_a} and \ref{fig:t1t2_ehningen_b} show the decoherence times $T_1$ and $T_2$, respectively, of the 27 qubits on {ibmq\_ehningen}. Qubits connected to two types of CX gates (Q-Bipotent) have a higher average value of $T_1$ and $T_2$ compared to qubits connected only to ECR-CX (Q-ECR) or only to direct-CX gates (Q-Direct). While the precise reason for this trend is unknown, it appears logical that more robust qubits are considered suitable for implementing both types of gates and are designated Q-Bipotent. This implies the advantage of using a combination of both types of CX gates.

Single-qubit gate error and readout error are shown in Figs.~\ref{fig:single_error_all_a} and \ref{fig:single_error_all_b}, respectively. The mean gate errors are comparable for three qubit types. However, the Q-ECR has a higher readout error overall than the Q-Direct and the Q-Bipotent. Throughout this paper, we use the default readout error mitigation provided by Qiskit \cite{qiskit} to reduce the impact of measurements on results. The averages of decoherence times and error rates of qubits are summarized in Table~\ref{table:single-qubit}.

The study on bipotent architecture shows that direct-CX gates have significantly reduced schedule duration. In addition, Q-Bipotent has the highest decoherence times $T_1$ and $T_2$ overall, while on average Q-Direct has a better $T_1$ than Q-ECR that has a better $T_2$.

\begin{figure*}[ht]
	\centering
	\includegraphics[width=\linewidth]{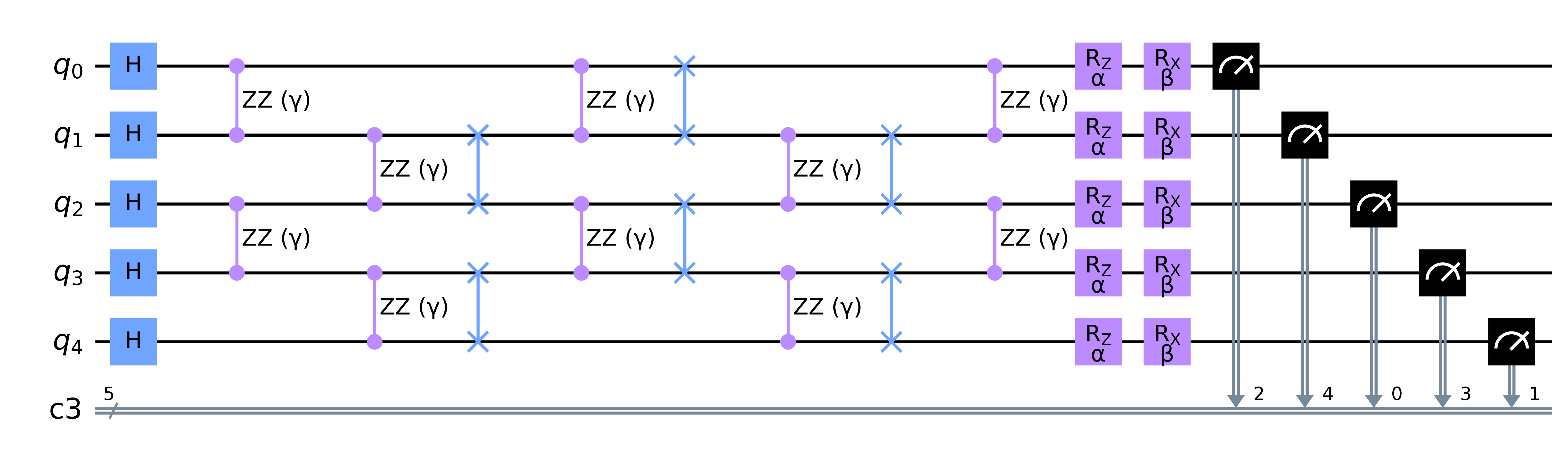}
	\vspace{-10pt}
	\caption{\label{fig:5q_transp} Pre-transpiled circuit of QAOA for PortOpt with 5 qubits and $p=1$ on a linear topology \cite{ji2022algorithm}. For higher $p$, the circuit between the initial state and measurement is repeated $p$ times, each time with new parameters.}
\end{figure*}

\section{Pulse-level Optimizations}
\label{sec:pulse-lev}

\subsection{Algorithm-Native Gate Set (ANGS)}
\label{subsec:ANGS}

Studying the properties of an algorithm is crucial to improve the performance of its implementation. One of the most important properties is the Algorithm-Native Gate Set (ANGS), which is the set of gates used to construct the algorithm. Efficient implementations of these gates will facilitate transpilation, enable exploitation of pulse-level optimization, and ultimately result in better performance.

Figure~\ref{fig:5q_transp} shows the pre-transpiled circuit of QAOA for PortOpt with 5 qubits and $p=1$ on a linear topology using the strategy in Ref.~\cite{ji2022algorithm} that provides optimal and scalable solutions. The resulting circuit has a structure similar to a swap network (see e.g., \cite{harrigan2021quantum}), but two layers are skipped per $p$, namely SWAP gates after the first and last layers of ZZ gates in every $p$. Thus, in the gate set \{H, RX, RZ, ZZ, ZZ-SWAP\}, the circuit with $n$ qubits has a depth of $2 + (n+2) \times p$ including initial state and measurement operators.

With the overall structure of QAOA circuit in Fig.~\ref{fig:5q_transp}, the ANGS of QAOA for PortOpt is \{H, ZZ, ZZ-SWAP, RZ,  RX\}, while that of QAOA for MaxCut is \{H, ZZ, ZZ-SWAP, RX\}, as no RZ is required for Eq.~\eqref{eq:cf_maxcut}. Therefore, we focus on improvements of these gate types based on different CX primitives.

\subsection{Analyzing two-qubit gates}

We now analyze two-qubit gates. The CZ, ZZ, and ZZ-SWAP gates are both undirected, which means that a more efficient implementation is possible with the native CX gate on the QPU. Additionally, gate cancellation at pulse-level provides opportunities for optimizations of gates implemented by ECR-CX gates.

\begin{figure*}[ht]
	\raggedright
	\includegraphics[width=0.40\linewidth]{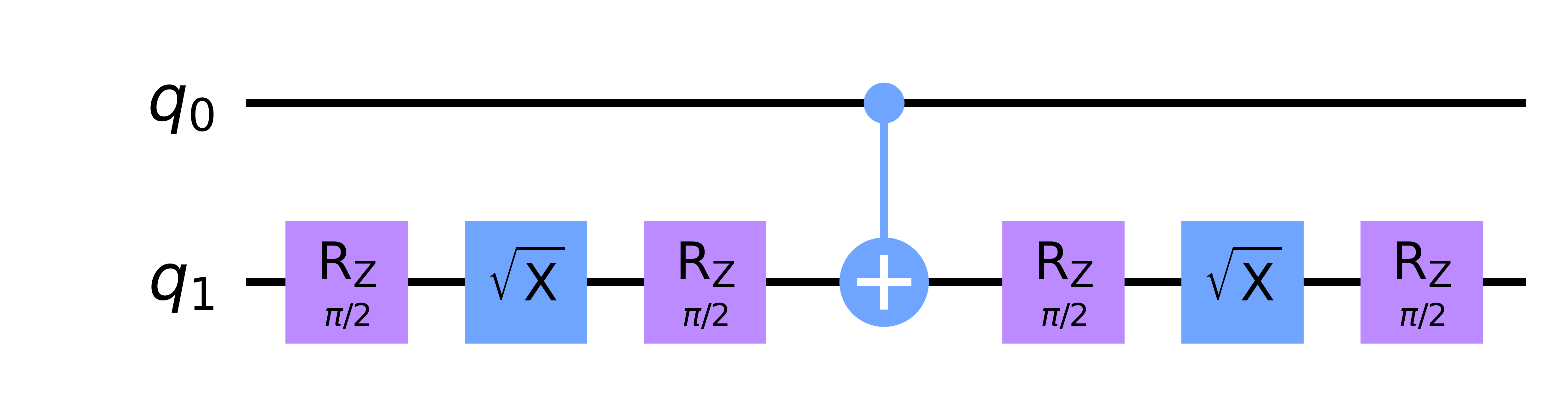}\hfill
	\put(-205,45){\textbf{(a)}}
	\phantomsubfloat{\label{fig:cz_gates_all_a}}
	\includegraphics[width=0.45\linewidth]{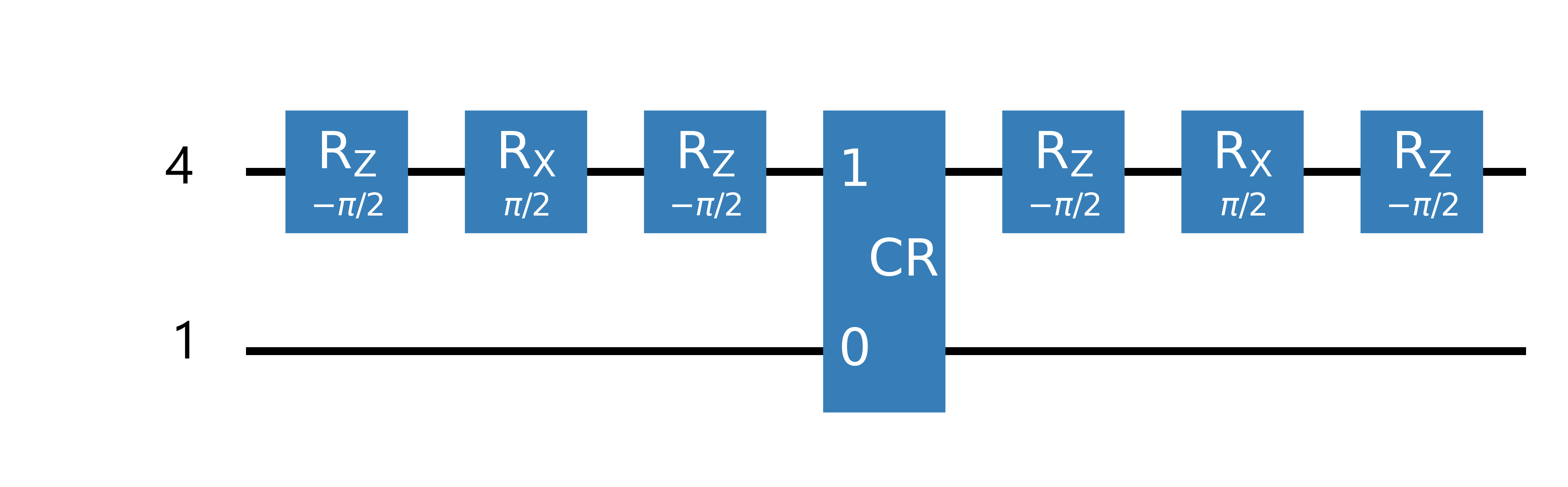}\hfill
	\put(-225,48){\textbf{(b)}}
	\phantomsubfloat{\label{fig:cz_gates_all_b}}
	\vspace{-12pt}
	
	\includegraphics[width=0.47\linewidth]{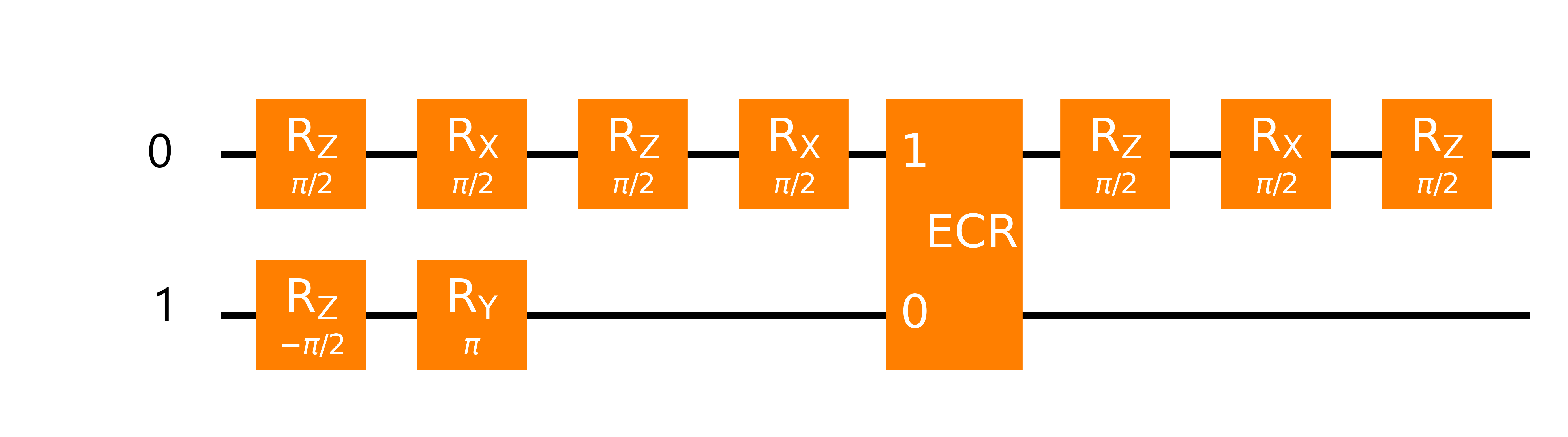}\hfill
	\put(-240,48){\textbf{(c)}}
	\phantomsubfloat{\label{fig:cz_gates_all_c}}
	\phantomsubfloat{\label{fig:cz_gates_all_d}}
	\includegraphics[width=0.33\linewidth]{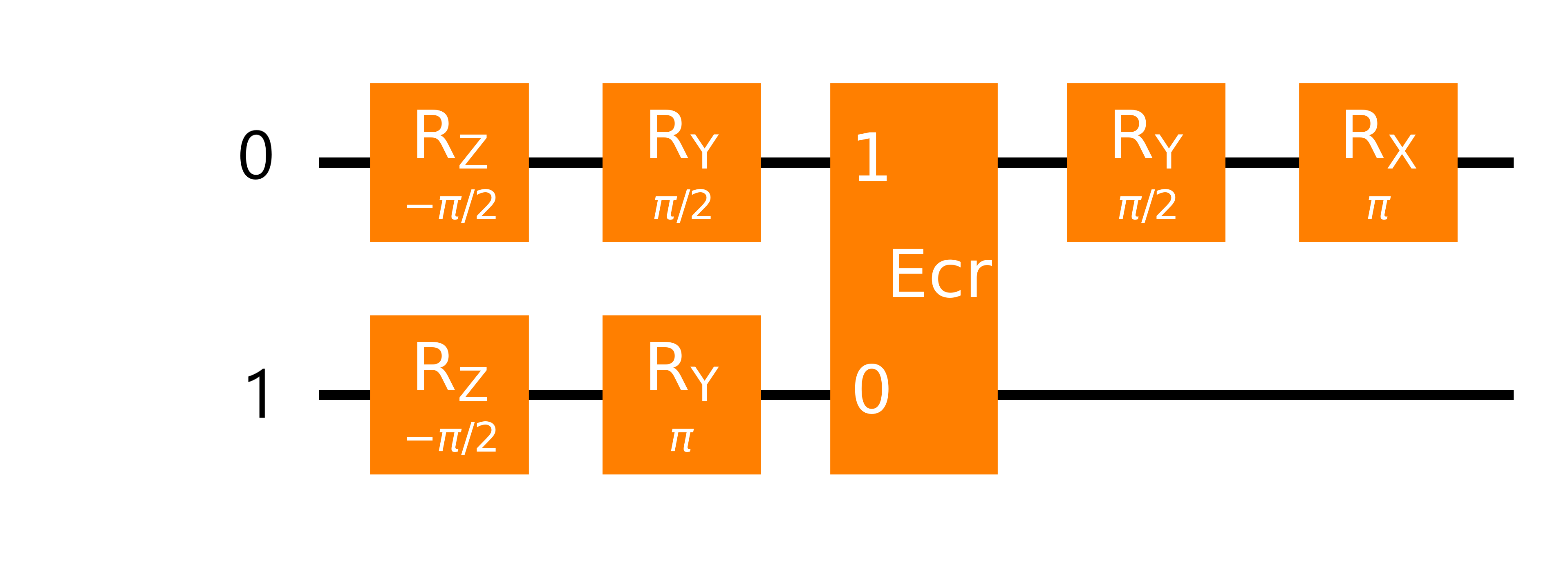}\hfill
	\put(-165,48){\textbf{(d)}}
	\vspace{-10pt}
	
	\includegraphics[width=0.25\linewidth]{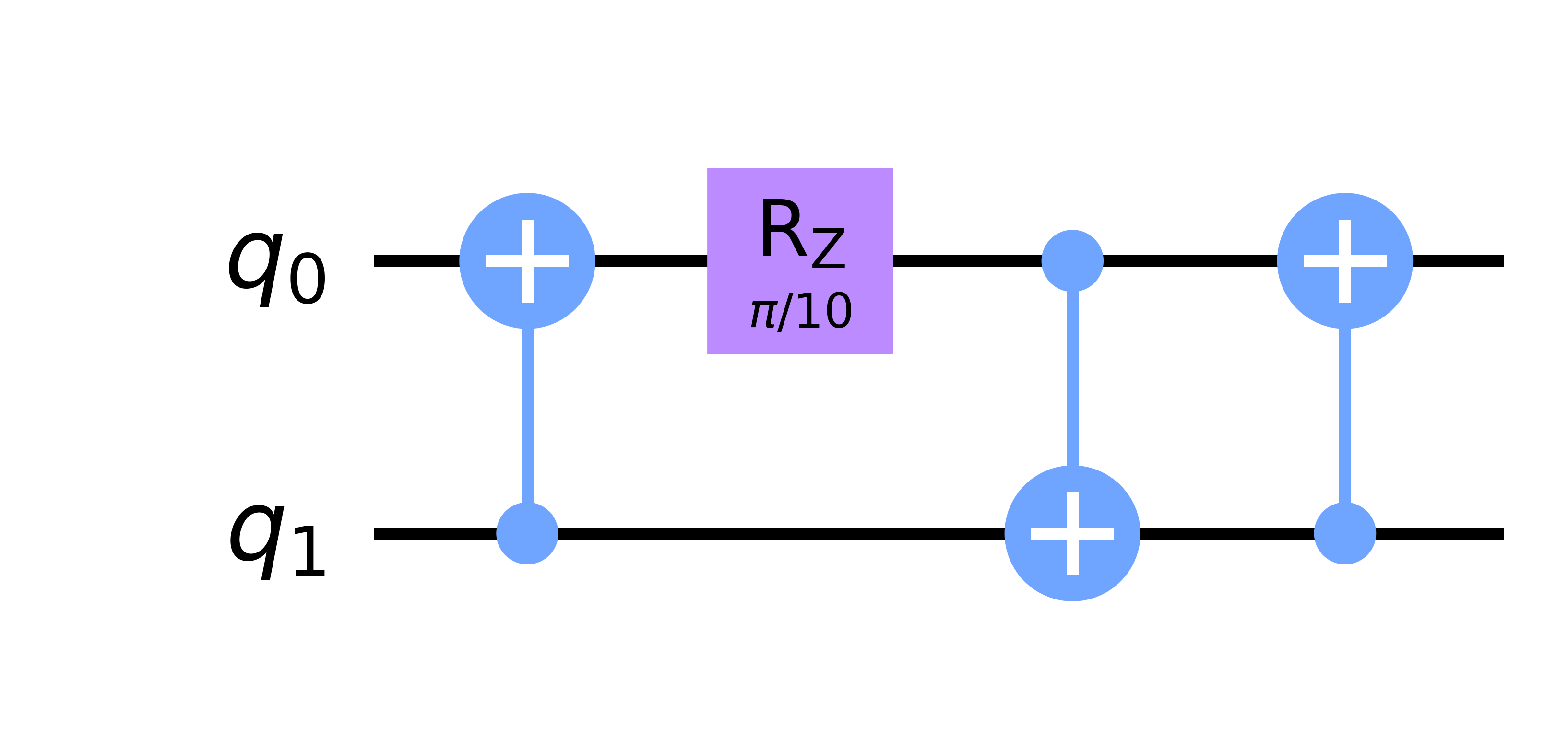}\hfill
	\put(-127,48){\textbf{(e)}}
	\phantomsubfloat{\label{fig:zzswap_gates_all_a}}
	\phantomsubfloat{\label{fig:zzswap_gates_all_b}}
	\includegraphics[width=0.45\linewidth]{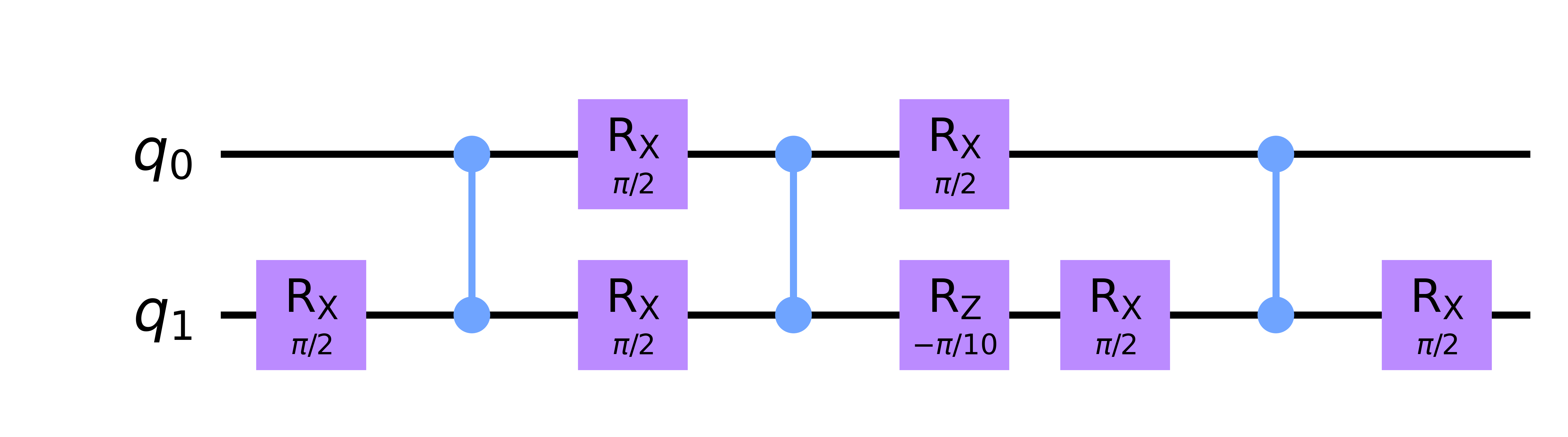}\hfill
	\put(-230,48){\textbf{(f)}}
	\vspace{-8pt}
	
	\phantomsubfloat{\label{fig:zzswap_gates_all_c}}
	\includegraphics[width=0.85\linewidth]{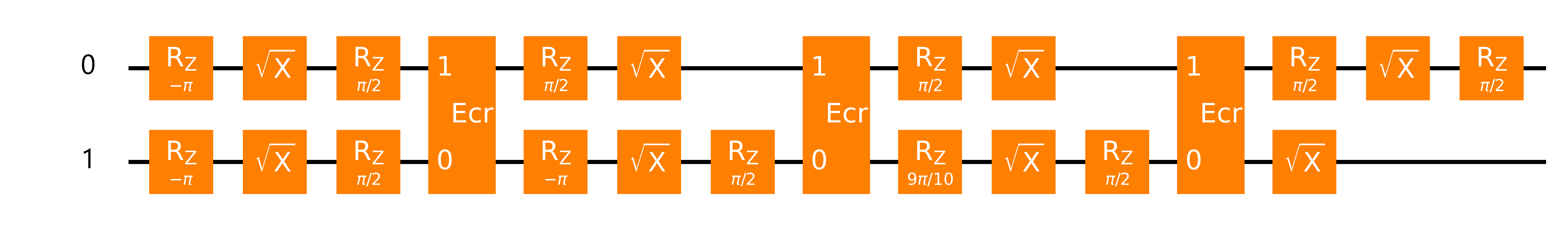}\hfill
	\put(-437,48){\textbf{(g)}}
	\caption{Different implementations (up to a global phase) of CZ and ZZ-SWAP gates on IBM's QPUs. (a) Decomposition of CZ gate. CZ gate implemented by (b) direct-CX gate on qubits [1, 4] and (c) ECR-CX gate on qubits [1, 0]. (d) CZ\textsubscript{OPT} gate, the pulse-level optimization of (c). (e) Decomposition of ZZ-SWAP gate. (f) Alternative decomposition of the ZZ-SWAP gate into CZ gates. (g) ZZ-SWAP\textsubscript{OPT} gate, the implementation of (f) on qubits [1, 0] using ECR-CX gates with the pulse-level optimization.}
\end{figure*}

Figure~\ref{fig:cz_gates_all_b} shows the default implementation of a CZ gate using direct-CX on qubits [1, 4] with a schedule duration of $309.3$\,ns. Although the implementation using ECR-CX on qubits [1, 0] shown in Fig.~\ref{fig:cz_gates_all_c} has a longer duration of $384$\,ns, it can be optimized at the pulse-level (CZ\textsubscript{OPT}) by reducing three single-qubit gates, resulting in a duration of $352$\,ns, as shown in Fig.~\ref{fig:cz_gates_all_d}.

The decompositions of ZZ-SWAP gate into CX gates and CZ gates \cite{hashim2022optimized} are shown in Figs.~\ref{fig:zzswap_gates_all_a} and \ref{fig:zzswap_gates_all_b}, respectively. In contrast to the CZ gate, the CX gate is directed. The ZZ-SWAP gate has a schedule duration of $992$\,ns on qubits [1, 0] and $800$\,ns on qubits [1, 4]. To implement the ZZ-SWAP gate with CZ gates, we use CZ\textsubscript{OPT} implemented by ECR-CX since the CZ gate is not a basis gate of IBM's QPUs, and then take advantage of single-qubit cancellation at the pulse-level, resulting in the same duration as the standard ZZ-SWAP gate. The optimized ZZ-SWAP gate (ZZ-SWAP\textsubscript{OPT}) is shown in Fig.~\ref{fig:zzswap_gates_all_c}.

The data show that implementing gates using vigorously optimized monolithic gates results in a shorter duration, while using ECR-CX gates allows for user-friendly pulse-level optimization without requiring additional calibration.

\begin{figure*}
	\centering
	\phantomsubfloat{\label{fig:qpt_rzz}}
	\includegraphics[width=0.485\linewidth]{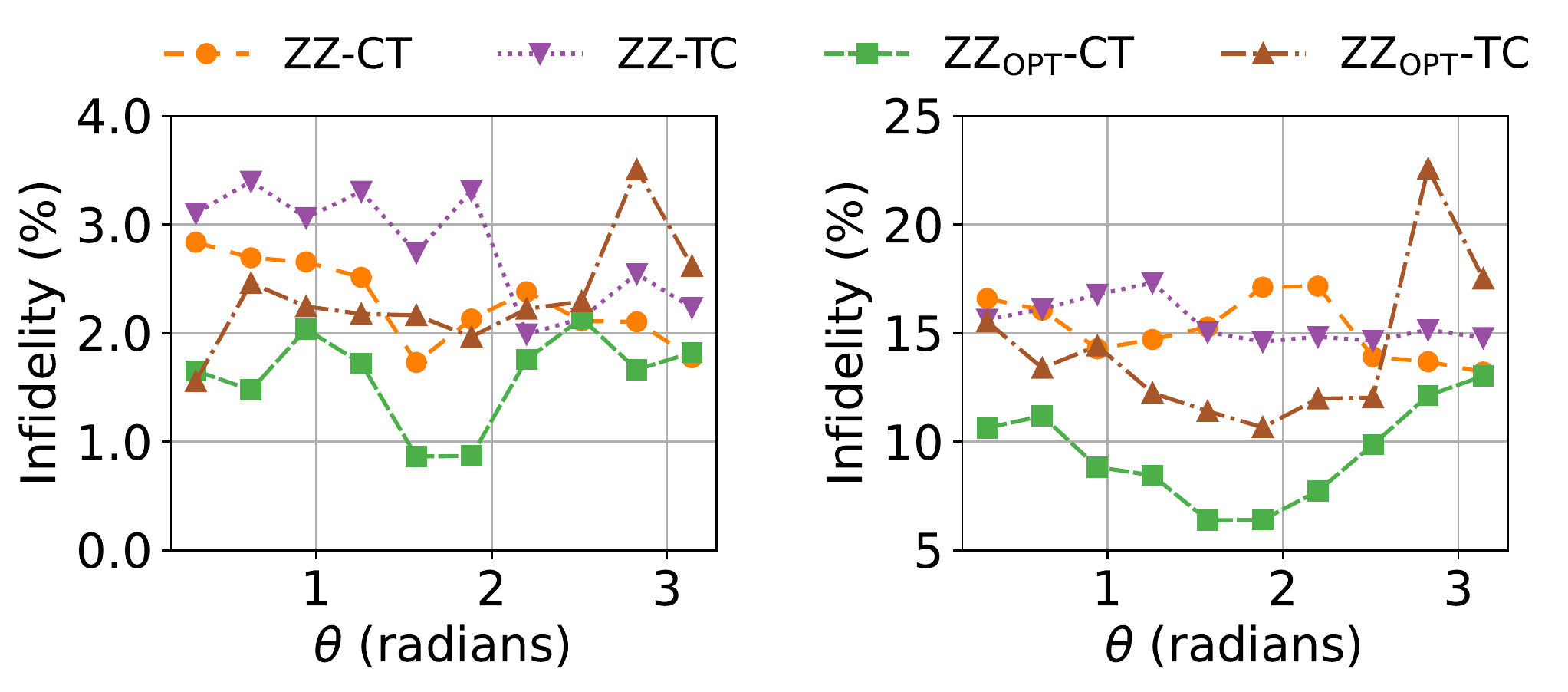}
	\put(-250,98){\textbf{(a)}}
	\phantomsubfloat{\label{fig:qpt_rzzswap}}
	\includegraphics[width=0.485\linewidth]{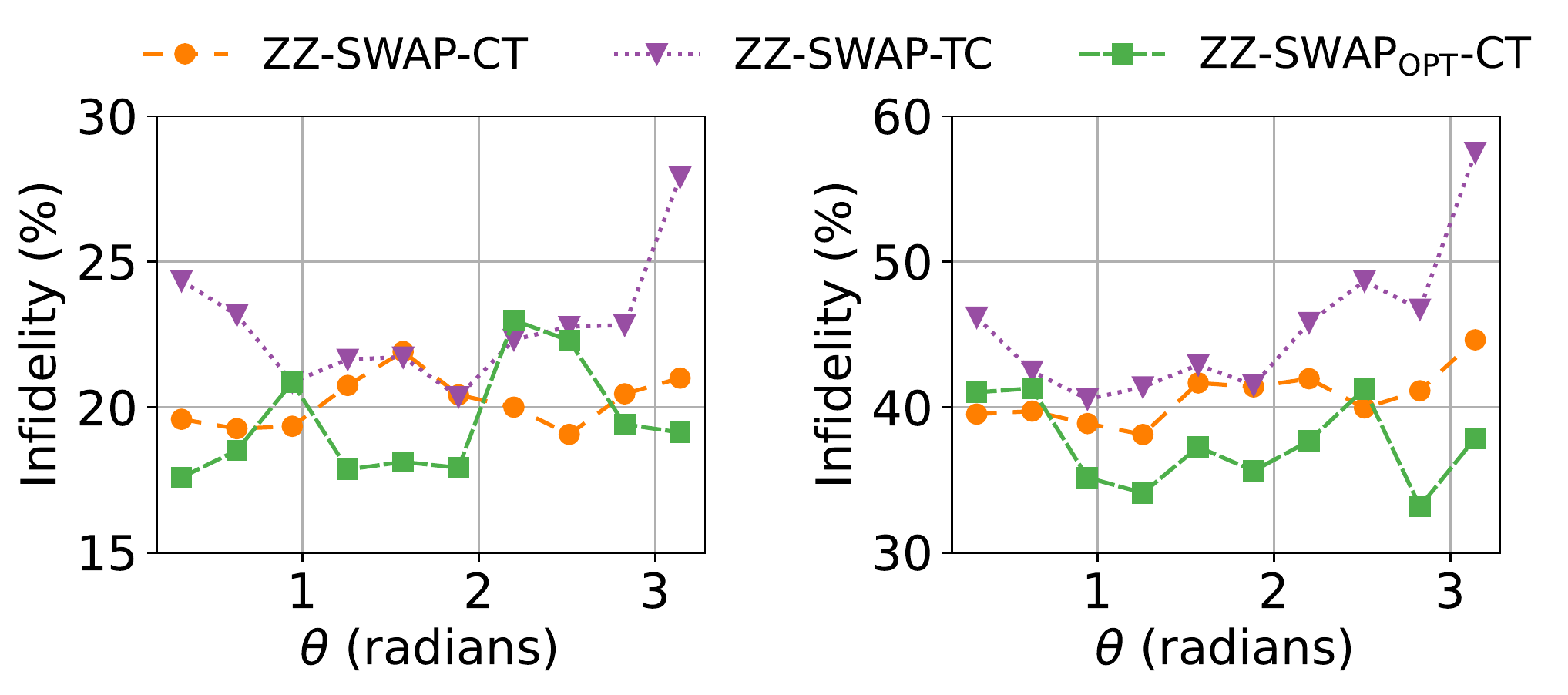}
	\put(-250,98){\textbf{(b)}}
	\caption{Infidelity determined by QPT on the {ibmq\_ehningen}. For (a), the ZZ gate acts on qubits [5, 8] with a gate repetition of 1 (left) and [1, 0] with a gate repetition of 10 (right). For (b), the ZZ-SWAP gate acts on qubits [1, 0] with a repeated number of gates of 10 (left) and 20 (right).}\label{fig:qpt_exp}
\end{figure*}

\subsection{Experimental results}

We have discussed the pulse-level optimization of gates implemented by ECR-CX gates. We now use the \emph{process fidelity} to benchmark the performance of default and optimized gates on the qubits that support ECR-CX gates. The \emph{process fidelity} can be determined by the Quantum Process Tomography (QPT) using Qiskit's standard implementation. The QPT is a method to characterize the actual behavior of quantum gates on QPUs. For QPT experiments, several initial states need to be prepared and then evolved, followed by numerous measurements at different measurement bases. The quantum channel is then reconstructed from the measurement data. The process fidelity between two quantum channels $\eta$ and $\xi$ is given by:
\begin{equation}
    F(\eta, \xi) = \Tr\left[\sqrt{\sqrt{\rho_\eta}\rho_\xi\sqrt{\rho_\eta}}\right]^2,
\end{equation}
where $\rho_\eta$ and $\rho_\xi$ are the normalized \emph{Choi} matrices for the channel $\eta$ and $\xi$, respectively. In our case, $\rho_\eta$ is the \emph{Choi} matrix of the ideal operator, while $\rho_\xi$ is the \emph{Choi} matrix determined by the QPT process on the QPU.

The ZZ and ZZ-SWAP denote gates implemented by the default basis gates of IBM's QPUs, while the ZZ\textsubscript{OPT} and ZZ-SWAP\textsubscript{OPT} refer to gates with pulse-level optimizations. 
In addition, we employ the notations CT and TC to distinguish between the gate implemented by CX gates with Control-Target (CT, in hardware native CX direction) and Target-Control (TC, opposite to hardware native CX direction).

Figure~\ref{fig:qpt_rzz} shows the infidelity $(1-F)$ of ZZ gate on qubits [5, 8] (left) with a gate repetition of 1 and on qubits [1, 0] (right) with a gate repetition of 10, respectively, as a function of angle. The schedule duration of ZZ\textsubscript{OPT}-CT and ZZ\textsubscript{OPT}-TC increases with angle, while that of ZZ-CT and ZZ-TC is constant since CX gate has a fixed duration and RZ gate is virtual. ZZ\textsubscript{OPT}-CT has the best overall performance as its duration is the shortest.

Figure~\ref{fig:qpt_rzzswap} shows the infidelity of the ZZ-SWAP gate on qubits [0, 1] with gate repetitions of 10 (left) and 20 (right), respectively. We benchmark the performance of ZZ-SWAP gate implemented by default CX gates considering CT and TC and by CZ\textsubscript{OPT} gates considering CT. The schedule duration of ZZ-SWAP-CT and ZZ-SWAP\textsubscript{OPT}-CT are the same, while that of ZZ-SWAP-TC is longer. The ZZ-SWAP\textsubscript{OPT} gate yields the best performance by decomposing the ZZ-SWAP into CZ gates, performing pulse-level optimization, and taking into account the Control-Target polarity of CX gates.

The results confirm that using hardware native polarity Control-Target to implement undirected gates, such as ZZ and ZZ-SWAP, is more efficient than using Target-Control polarity which results in a longer duration. In addition, gates optimized at pulse-level perform better than their default implementation. In the benchmarking below, we implement the undirected gates with Control-Target.

\begin{figure*}[ht]
	\includegraphics[width=\linewidth]{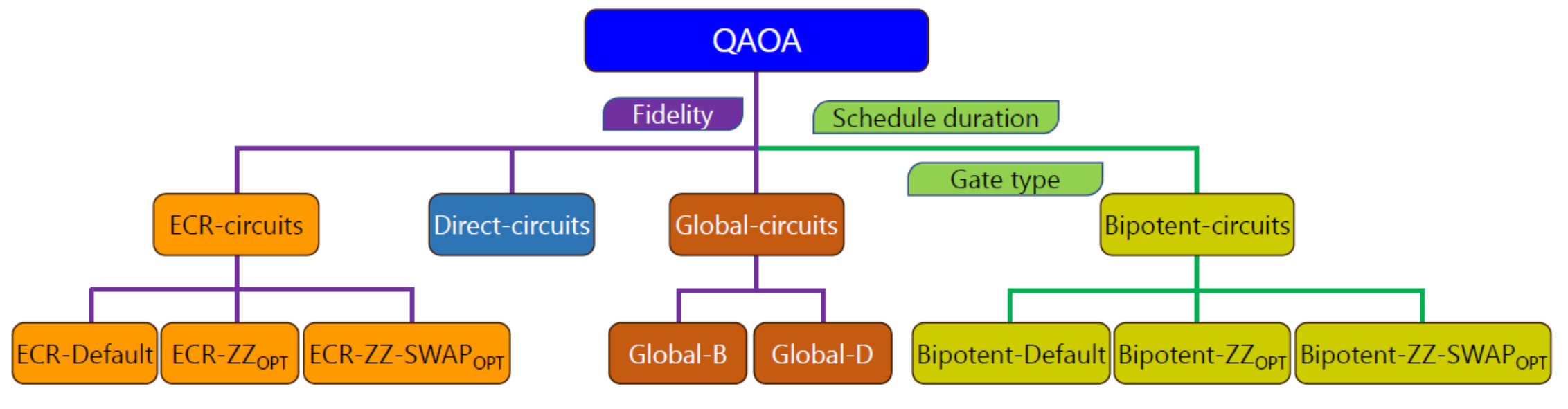}\hfill
	\caption{QAOA circuits. ECR-circuits, Direct-circuits, and Global-circuits consider fidelity and restrict the CX gate type in the circuit to only ECR-CX, only direct-CX, and regardless of its type, respectively, whereas Bipotent-circuits have a trade-off between fidelity, schedule duration, and gate type.}\label{fig:qaoa_circuits}
\end{figure*}

\section{Benchmarking with QAOA}
\label{sec:bench}

\subsection{Benchmarks and performance metrics}

We consider as benchmarks the circuits with different numbers of qubits and depths $p$. The maximum number of qubits that can be connected linearly using only ECR-CX gates is five, e.g., [9, 8, 11, 14, 16], as illustrated in Fig.~\ref{fig:gate_map_ehningen}. Therefore, we limited the number of qubits to five to allow a full comparison of qubits with different CX gate types.

The five-asset PortOpt instance that runs on five qubits (5Q) is taken from \cite{brandhofer2023benchmarking}, and its first three and four assets are used for the three- (3Q) and four-qubit (4Q) instances, respectively. The values of the variables $q$, $B$, $A$, and $\lambda$ vary across the three cases considered. For 3Q, they are $0.33$, $2$, $0$, and $20.97$, respectively. For 4Q, they are $0.33$, $2$, $0.13$, and $17.99$, respectively. Finally, for 5Q, they are $0.33$, $3$, $0.07$, and $17.51$, respectively. For MaxCut we consider the \emph{complete graph}, which results in the same connectivity requirements as the dense PortOpt problem instances.

The AR and SP defined in Sec. \ref{sec:background-qaoa} are used to evaluate the performance. We simulate the circuits in the noiseless case with the QASM-simulator to get the values of AR and SP of QAOA for PortOpt and MaxCut that can be used as a baseline. For the demonstrations on QPUs, we set the number of shots to 50000 for each circuit. Table \ref{table:protopt_arsp} presents the simulated values of the AR and SP of QAOA for PortOpt with $p\in\{1, 2, ..., 5\}$, while Table \ref{table:maxcut_arsp} displays the results of QAOA for MaxCut. The data show that QAOA with 5 qubits for PortOpt exhibits smaller AR and SP values compared to MaxCut for the specific instances considered.
\begin{table}[b]
\begin{ruledtabular}
  \centering
  \caption{AR and SP of QAOA for PortOpt without noise.}
  \label{table:protopt_arsp}
  \begin{tabular}{llllll}
    \textbf{$p$} & \textbf{1} & \textbf{2}&\textbf{3} & \textbf{4} & \textbf{5}\\
    3Q-AR & 0.993 & 0.996 & 0.998 & 1.000 & 1.000 \\
    3Q-SP (\%) & 99.34 & 99.57 & 99.83 & 99.99 & 99.98 \\
    4Q-AR & 0.525 & 0.674 & 0.892 & 0.930 & 0.935 \\
    4Q-SP (\%) & 21.18 & 29.04 & 44.30 & 45.61 & 46.97 \\
    5Q-AR & 0.513 & 0.819 & 0.828 & 0.898 & 0.909 \\
    5Q-SP (\%) & 17.70 & 48.61 & 45.28 & 55.33 & 58.35 \\
  \end{tabular}
\end{ruledtabular}


\begin{ruledtabular}
  \centering
  \caption{AR and SP of QAOA for MaxCut without noise.}
  \label{table:maxcut_arsp}
  \begin{tabular}{llllll}
    \textbf{$p$} & \textbf{1} & \textbf{2}&\textbf{3} & \textbf{4} & \textbf{5}\\
    5Q-AR & 0.914 & 0.972 & 0.996 & 0.997 & 0.997  \\
    5Q-SP (\%) & 87.79 & 96.99 & 99.45 & 99.62 & 99.72 \\
  \end{tabular}
\end{ruledtabular}
\end{table}

In addition to AR and SP, we use the schedule duration and the number of CX gates to study the properties of the circuit. A lower schedule duration means that the circuit executes faster, while a reduced number of CX gates implies pulse-level optimizations.

\subsection{Methodology}

To implement QAOA, we first prepare pre-transpiled circuits as described in Sec. \ref{subsec:ANGS}. This process ensures that the circuits satisfy the connectivity constraints on a linear topology. The next crucial step is to identify high-quality qubits to execute them. To evaluate the qubit quality, we first consider only the circuit fidelity \cite{ji2022calibration}. As illustrated in Fig.~\ref{fig:qaoa_circuits}, we consider four families of QAOA circuits for evaluation. ECR-, Direct- and Global-circuits connected with violet (dark gray) lines use the qubits selected by Mapomatic \cite{mapomatic} aiming to maximize the fidelity of the circuit. In comparison, the Bipotent-circuits connected with green (light gray) lines take into account the schedule duration, fidelity, and gate type. Moreover, ECR-circuits utilize only the ECR-CX gates (with or without pulse-level optimizations). Direct-circuits are composed of only direct-CX gates, while Global-circuits and Bipotent-circuits contain mixtures of both gate types.

To construct an ECR-circuit with $k$ qubits, Mapomatic \cite{mapomatic} selects a maximum-fidelity linear arrangement of $k$ qubits such that ECR-CX gates are available on neighboring qubits. We use three versions of such circuits: ECR-Default that employs default implementations of ZZ and ZZ-SWAP gates; ECR-ZZ\textsubscript{OPT} that utilizes pulse-level-optimized ZZ gate performed with the reduced CR gate; and ECR-ZZ-SWAP\textsubscript{OPT} that leverages pulse-level optimizations for ZZ gates and reconstructs ZZ-SWAP gates with CZ\textsubscript{OPT} gates. We maintain the same set of qubits for each data point if optimizations are implemented.

Direct-circuits exclusively utilize direct-CX gates, and best-fidelity linear arrangements of qubits that support such gates are selected, again using Mapomatic. It is worth noting that there are no pulse-level optimizations in Direct-circuits. Global-circuits are best-fidelity linear arrangements of arbitrary qubits, regardless of whether their connections support ECR-CX or direct-CX. It is possible, but not necessary, that all selected qubits of a Global-circuit have only direct-CX links. We call such configurations Global-D (for ``direct''), and configurations with mixtures of direct-CX and ECR-CX Global-B (for ``bipotent''). We observe that due to the higher error rate of ECR-CX gates, as shown in Fig.~\ref{fig:cx_all_ehningen}, only ECR-CX-linked qubits are not selected in Global-circuits.

In addition to those circuits where qubits are selected based on the circuit fidelity, we consider a further bipotent arrangement, called Bipotent-circuits. As the schedule duration of a circuit on the selected qubits has a fixed value and can be determined directly by Qiskit, Bipotent-circuit optimizes duration while controlling its fidelity. We focus only on qubits with below-average single-qubit and two-qubit error rates. Among these qubits, we select a linear arrangement on which the pulse-level optimized circuit has the shortest schedule duration. Moreover, we enforce bipotency by requiring at least one ECR-CX and at least one direct-CX gate. Pulse-level optimizations can be applied to ECR-CX gates within such circuits; the resulting circuits are labeled Bipotent-Default, Bipotent-ZZ\textsubscript{OPT} and Bipotent-ZZ-SWAP\textsubscript{OPT}, analogously to ECR-circuits above.

\subsection{Benchmarking results}
\label{subsec:bench}

\subsubsection{QAOA for PortOpt}
\label{subsubsec:bench_qaoa_portopt}

In this section, we first benchmark the performance of QAOA for PortOpt with ECR-circuits, Global-circuits, and Direct-circuits. Then, we demonstrate the performance of QAOA with Bipotent-circuits and explore the pulse-level optimizations.

\begin{figure*}[ht]
	\centering
	\includegraphics[width=\linewidth]{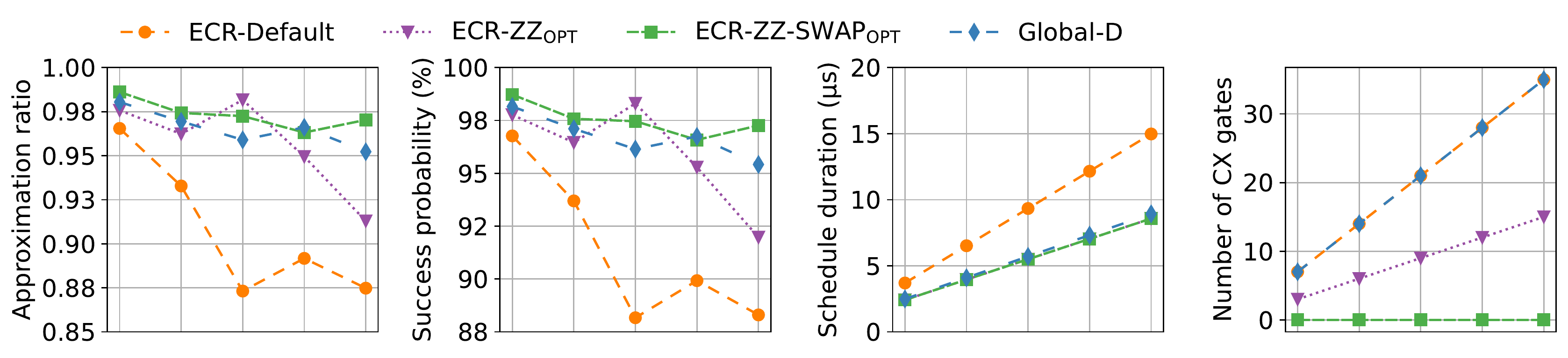}\hfill
	\put(-520,105){\textbf{(a)}}
	\phantomsubfloat{\label{fig:portopt_ecr_a}}
	\vspace{-30pt}
	
	\phantomsubfloat{\label{fig:portopt_ecr_b}}
	\includegraphics[width=\linewidth]{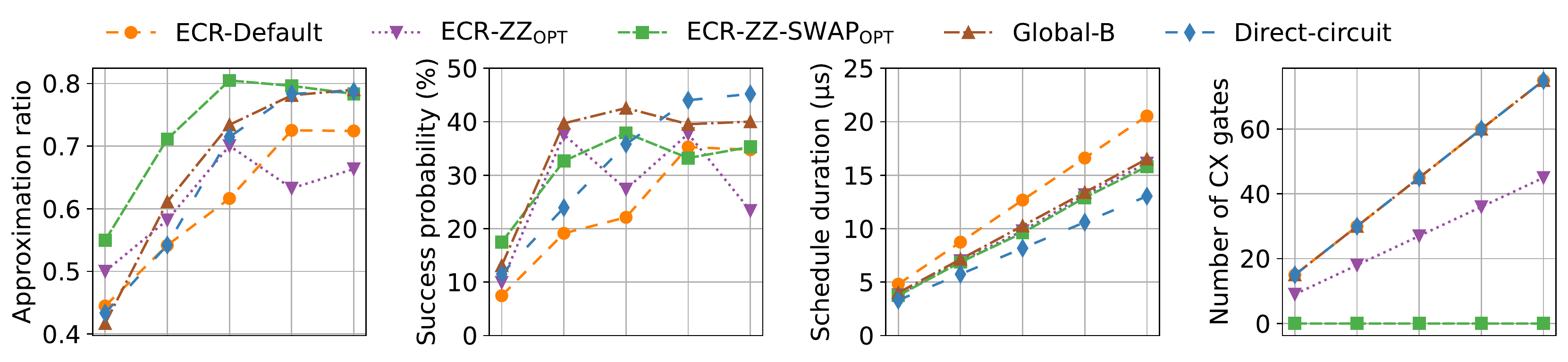}\hfill
	\put(-520,105){\textbf{(b)}}
	\vspace{-20pt}
	
	\phantomsubfloat{\label{fig:portopt_ecr_c}}
	\includegraphics[width=\linewidth]{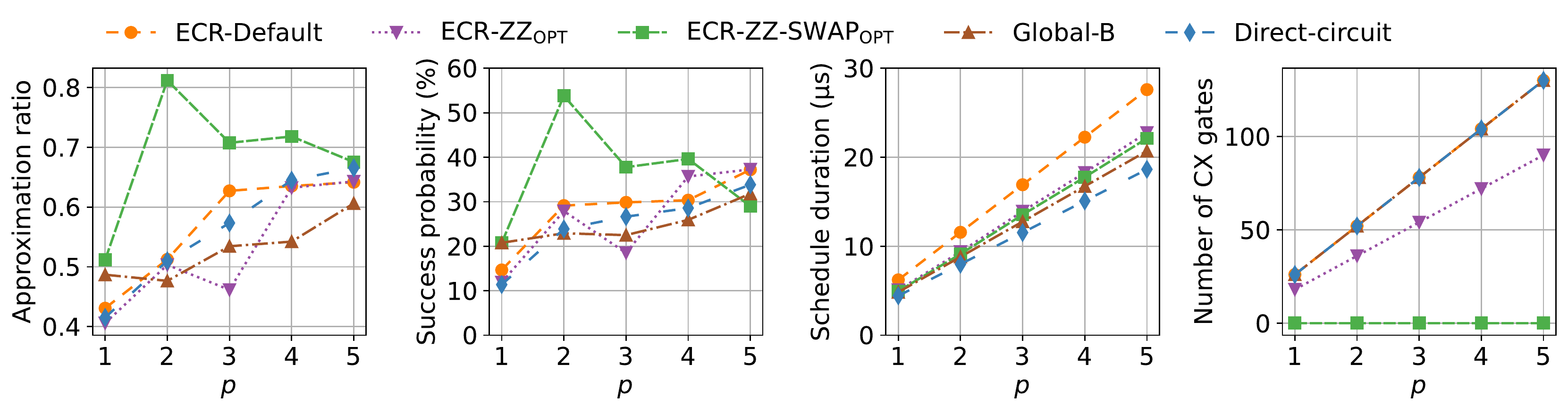}\hfill
	\put(-520,122){\textbf{(c)}}
	\caption{Approximation ratio, success probability, schedule duration, and CX gate count of QAOA for PortOpt with $p$ from 1 to 5 using ECR-, Global-, and Direct-circuits on {ibmq\_ehningen} for (a) 3Q, (b) 4Q, and (c) 5Q. The average CX gate error rates are $0.48\%$ in all Global-circuits, they are $0.78\%$, $0.87\%$, and $0.89\%$ in ECR-circuits for 3Q, 4Q, and 5Q, respectively, while $0.48\%$ and $0.60\%$ in Direct-circuits for 4Q and 5Q, respectively.}\label{fig:portopt_ecr}
\end{figure*}

\begin{figure*}[ht]
	\centering
	\includegraphics[width=\textwidth]{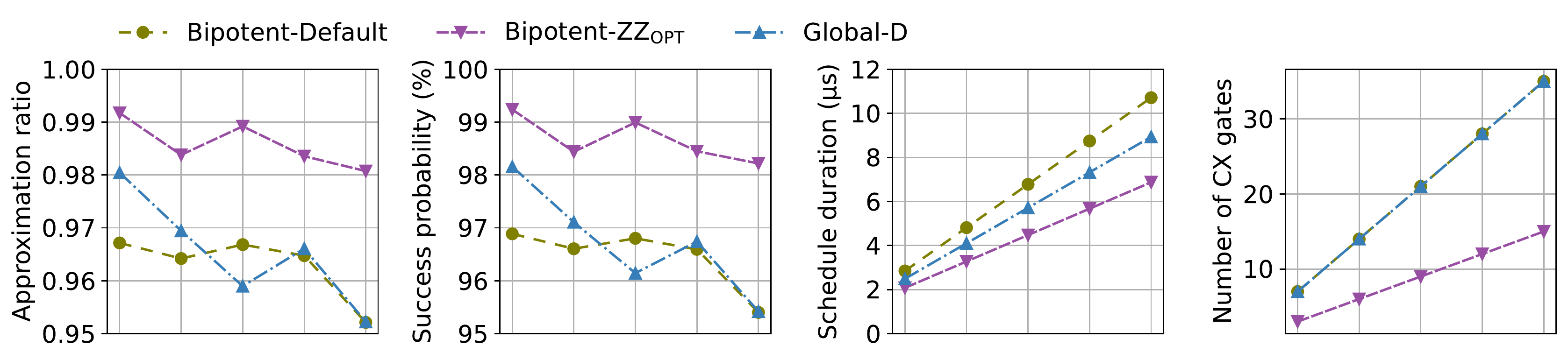}\hfill
	\put(-520,105){\textbf{(a)}}
	\phantomsubfloat{\label{fig:portopt_hybrid_a}}
	\vspace{-30pt}
	
	\phantomsubfloat{\label{fig:portopt_hybrid_b}}
	\includegraphics[width=\textwidth]{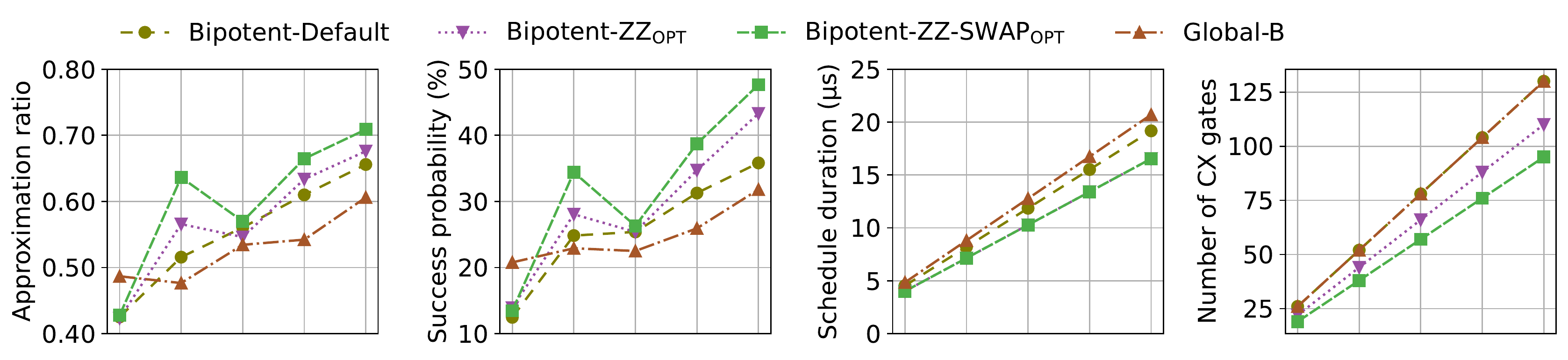}\hfill
	\put(-520,105){\textbf{(b)}}
	\vspace{-20pt}
	
	\phantomsubfloat{\label{fig:portopt_hybrid_c}}
	\includegraphics[width=\textwidth]{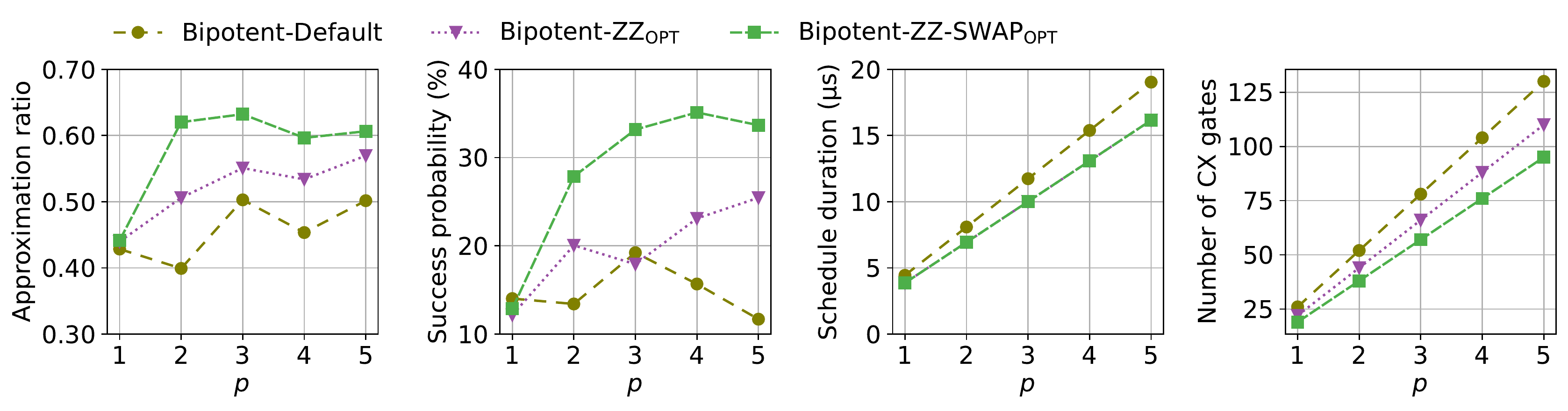}\hfill
	\put(-520,120){\textbf{(c)}}
	
	\caption{Benchmarking results of QAOA for PortOpt using Bipotent- and Global-circuits on {ibmq\_ehningen} for (a) 3Q, (b) 5Q, and (c) on {ibm\_auckland} for 5Q. The average CX gate error rates for each data point in (a), (b), and (c) are 0.60\%, 0.65\%, and 0.67\%, respectively.
	}\label{fig:portopt_hybrid}
\end{figure*}

The results of QAOA-PortOpt with 3Q, 4Q, and 5Q are shown in Figs.~\ref{fig:portopt_ecr_a}, \ref{fig:portopt_ecr_b}, and \ref{fig:portopt_ecr_c}, respectively. The ECR-Default has the largest schedule duration. In the ECR-ZZ\textsubscript{OPT}, the number of CX gates is reduced as the ZZ gates are optimized. Furthermore, the ZZ-SWAP\textsubscript{OPT} gate is implemented by CZ\textsubscript{OPT} gates, therefore there are no CX gates in the ECR-ZZ-SWAP\textsubscript{OPT}.

In Fig.~\ref{fig:portopt_ecr_a}, the qubits used in the ECR-circuits are [16, 14, 11], while in Global-D they are [18, 21, 23]. The ECR-ZZ-SWAP\textsubscript{OPT} has the best performance, although the average CX gate error rate in ECR-circuits is 1.6 times higher than that in the Global-D. Optimizing both ZZ and ZZ-SWAP gates yields better and more stable AR and SP than optimizing only ZZ gates.

Figure~\ref{fig:portopt_ecr_b} shows the results of QAOA using ECR-circuits with qubits [16, 14, 11, 8], Global-B with qubits [18, 21, 23, 24], and Direct-circuit with qubits [1, 4, 7, 6]. Despite having the highest fidelity, Global-B does not perform the best. In comparison, ECR-ZZ-SWAP\textsubscript{OPT} achieves the highest AR values, although its average CX gate error rate is 1.8 times higher and the duration is much longer. We believe this advantage comes from the combination of ZZ\textsubscript{OPT} and ZZ-SWAP\textsubscript{OPT} gates. The Direct-circuit has a shorter schedule duration than Global-B resulting in comparable AR values and better SP values for higher $p$.

Figure~\ref{fig:portopt_ecr_c} shows a performance comparison of 5Q-QAOA for PortOpt using ECR-circuits with qubits [16, 14, 11, 8, 9], Global-B with qubits [17, 18, 21, 23, 24] including two ECR-CX gates, and Direct-circuit with qubits [6, 7, 4, 1, 2]. While the Global-B and Direct-circuit exhibit better fidelity and shorter schedule durations, ECR-ZZ-SWAP\textsubscript{OPT} outperforms them with the highest AR values. Therefore, despite its longer duration, ECR-ZZ-SWAP\textsubscript{OPT} can be considered as a promising alternative for achieving higher accuracy in quantum computing. A possible explanation is that combining ZZ\textsubscript{OPT} and ZZ-SWAP\textsubscript{OPT} produces a pulse that is more resilient to noise than the others.

\begin{figure}
	\centering
	\vspace{-5pt}
	\includegraphics[width=\columnwidth]{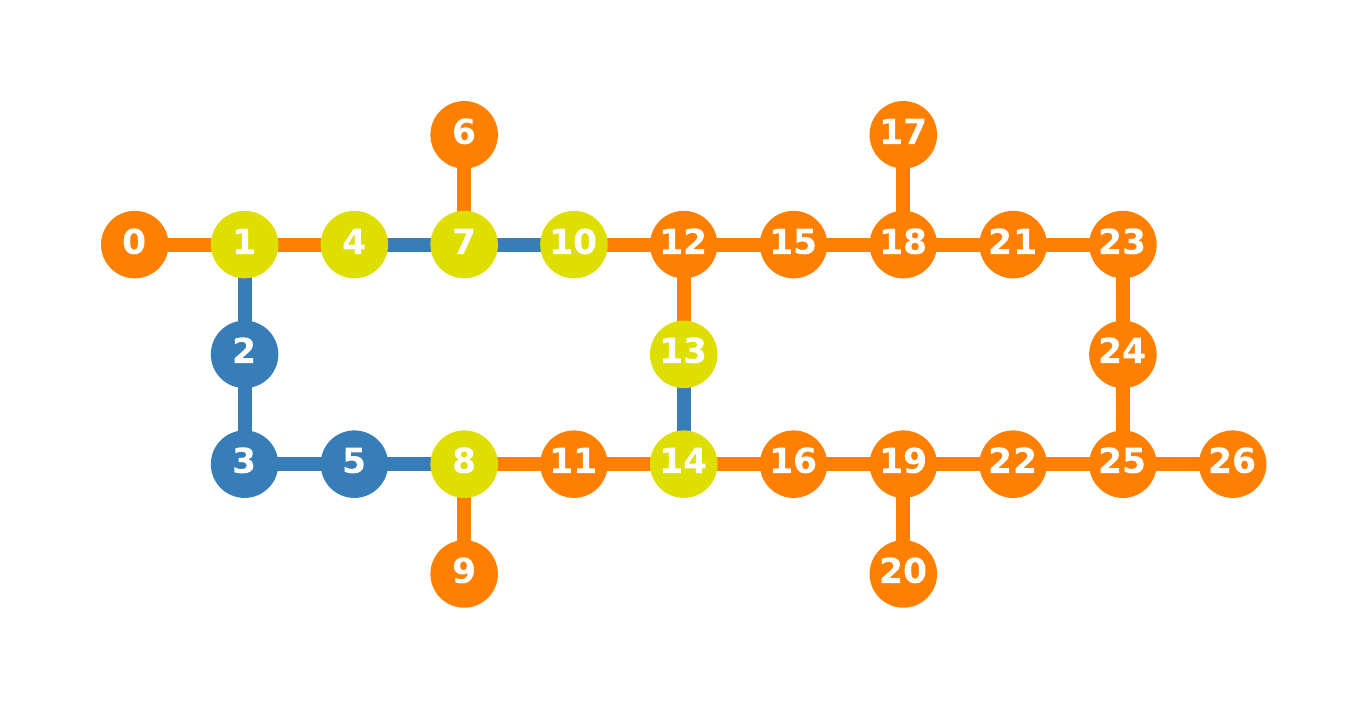}
	\vspace{-15pt}
	\caption{Bipotent architecture of {ibm\_auckland}. It has the same label as {ibmq\_ehningen}.}\label{fig:gate_map_auckland}
\end{figure}

We now benchmark the performance of QAOA with Bipotent-circuits (optimized for schedule duration rather than for fidelity). Figure~\ref{fig:portopt_hybrid_a} shows the results of 3Q-QAOA on {ibmq\_ehningen}. The qubits used in Bipotent-circuits are [22, 25, 26] including one ECR-CX on qubits [22, 25] and one direct-CX on qubits [25, 26]. Compared to Bipotent-Default and Global-D, Bipotent-ZZ\textsubscript{OPT} demonstrates better performance, as indicated by its higher AR and SP values. The decrease in the number of CX gates achieved through the use of optimized ZZ gates suggests that partial ZZ gates originally implemented with default CX gates have been optimized at the pulse-level. This optimization leads to a reduction in schedule duration.

The Bipotent-circuits in Fig.~\ref{fig:portopt_hybrid_b} employ a set of five qubits, namely [11, 14, 13, 12, 10], among which only a single pair of qubits is capable of performing an ECR-CX gate. The optimization of ZZ gates implemented using ECR-CX gates in Bipotent-ZZ\textsubscript{OPT} leads to a reduction in the number of CX gates and a shorter schedule duration. Bipotent-ZZ-SWAP\textsubscript{OPT} further optimizes ZZ-SWAP gates by replacing certain ECR-CX-based ZZ-SWAP gates with CZ\textsubscript{OPT}-based ZZ-SWAP\textsubscript{OPT} gates, thus further decreasing the number of CX gates. Although Global-B has better fidelity, the demonstrations show that Bipotent-circuits outperform it for $p \ge 2$. Overall, among all the methods tested, Bipotent-ZZ-SWAP\textsubscript{OPT} exhibits the best performance in terms of the improvement in both the AR and SP metrics, making it a promising approach for quantum circuit optimization.

Figure~\ref{fig:portopt_hybrid_c} presents the benchmark results of 5Q-QAOA using qubits [11, 8, 5, 3, 2] on {ibm\_auckland}. It is worth noting that {ibm\_auckland} has a different bipotent architecture compared to {ibmq\_ehningen}, as illustrated in Fig.~\ref{fig:gate_map_auckland}. The results on {ibm\_auckland} are consistent with the trend of 5Q-QAOA for PortOpt observed on {ibmq\_ehningen}. Specifically, we observe that QAOA with Bipotent-ZZ-SWAP\textsubscript{OPT} outperforms the others. These results highlight the potential of the proposed optimization approach in improving the performance of quantum algorithms on other QPUs.

The data show that leveraging a combination of ZZ\textsubscript{OPT} and ZZ-SWAP\textsubscript{OPT} gates can significantly enhance both AR and SP, surpassing the performance achieved by solely optimizing ZZ gates. This approach also outperforms Global-circuits with the highest fidelity. Specifically, when compared to Global-circuits, we observe an improvement in AR of up to 70\%, on average 29\%, for ECR-ZZ-SWAP\textsubscript{OPT} at $p = 2$.  Similarly, on ibm\_auckland, Bipotent-ZZ-SWAP\textsubscript{OPT} shows an improvement of up to 55\%, on average 27\%, compared to Bipotent-Default. On ibmq\_ehningen, Bipotent-ZZ-SWAP\textsubscript{OPT} achieves an improvement of up to 34\%, on average 14\%, compared to Global-B.

\subsubsection{QAOA for MaxCut}
\label{subsubsec:bench_qaoa_maxcut}

\begin{figure*}
	\centering
	\includegraphics[width=\textwidth]{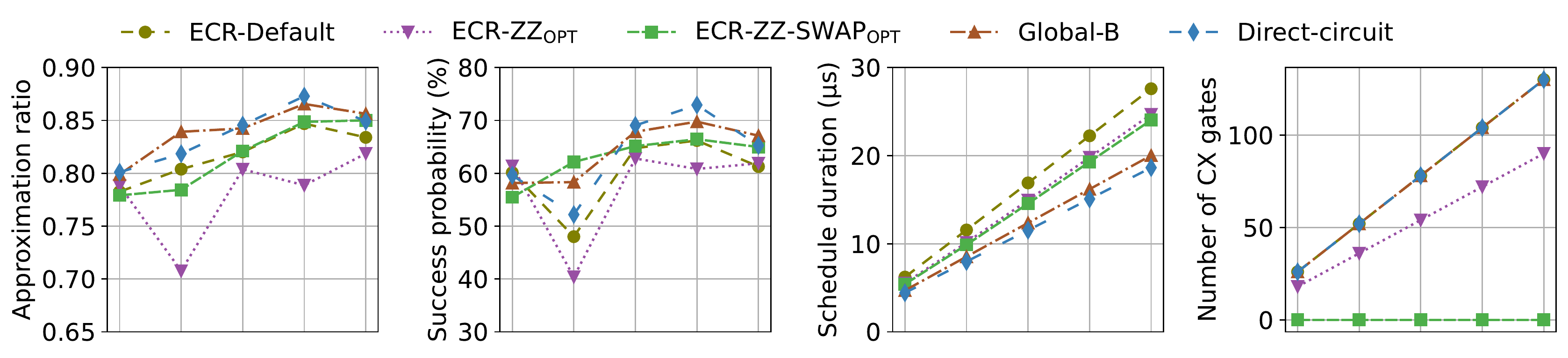}
	\put(-520,105){\textbf{(a)}}
	\phantomsubfloat{\label{fig:maxcut_ecr}}
	\vspace{-28pt}
	
	\phantomsubfloat{\label{fig:maxcut_hybrid}}
	\includegraphics[width=\textwidth]{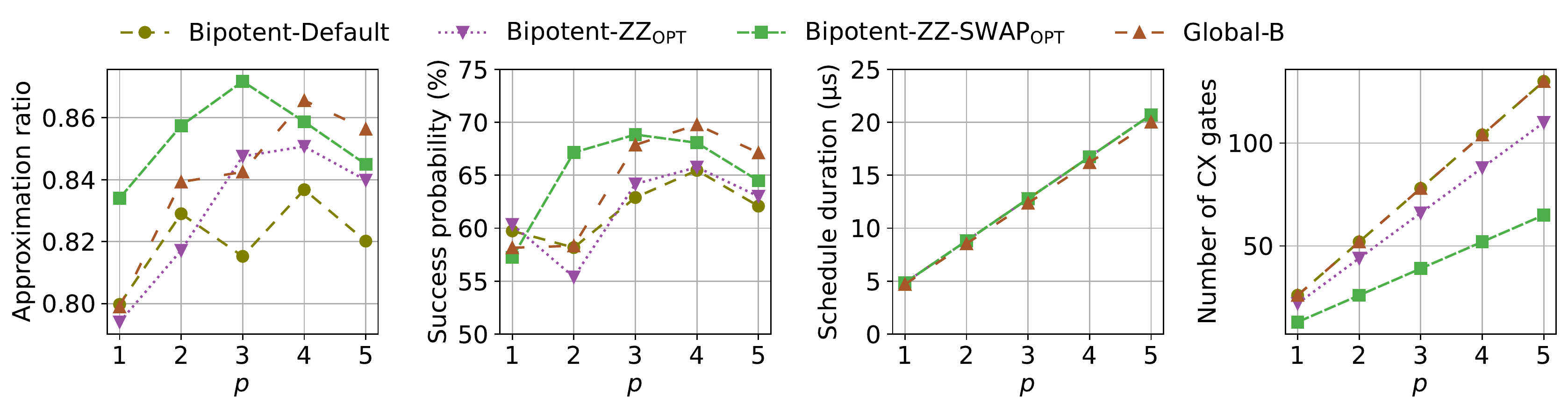}
	\put(-520,120){\textbf{(b)}}
	\caption{Benchmarking results of 5Q-QAOA for MaxCut on {ibmq\_ehningen} using two sets of circuits: (a) ECR-, Global-, and Direct-circuits, and (b) Bipotent- and Global-circuits. The average CX gate error rates in the ECR-circuits, Global-B, and Direct-circuit are 0.89\%, 0.48\%, and 0.60\%, respectively, while that of each data point in (b) is 0.48\%.}\label{fig:maxcut_res}
\end{figure*}

We now investigate the performance of QAOA for MaxCut with 5 qubits. The qubits used in ECR-circuits, Global-B, and Direct-circuit are [16, 14, 11, 8, 9], [6, 7, 4, 1, 0], and [6, 7, 4, 1, 2], respectively. Figure~\ref{fig:maxcut_ecr} shows that Global-B, which has the highest fidelity, performs better for smaller $p$, whereas Direct-circuit, which has the lowest schedule duration, performs better for larger $p$. These results suggest that, when using default IBM gates, fidelity is more critical for medium-scale circuits, while a shorter schedule duration is more important for larger circuits. One possible explanation is that, as the size of the circuit increases, the effects of decoherence become dominant and prevail over the variability in qubit error rates. ECR-circuits with higher error rates and longer schedule durations lead to lower AR compared to other methods. However, by combining ZZ\textsubscript{OPT} and ZZ-SWAP\textsubscript{OPT}, ECR-ZZ-SWAP\textsubscript{OPT} achieves stable performance in both AR and SP.

Figure~\ref{fig:maxcut_hybrid} shows the results of QAOA using Bipotent-circuits with qubits [17, 18, 21, 23, 24] including two ECR-CX and two direct-CX gates. While all three types of Bipotent-circuits have the same schedule duration, Bipotent-ZZ-SWAP\textsubscript{OPT} exhibits a clear advantage, as demonstrated by significant improvements in both AR and SP. However, Global-B exhibits better performance for larger values of $p$ due to its shorter duration.

Our study highlights the importance of efficient gate implementation, achieved through increasing fidelity and reducing schedule duration, in enhancing the algorithm's performance. However, we found that the use of different gate types in a circuit can result in significantly divergent performances, even when the duration is identical. Therefore, careful consideration must be given to the selection of gate types to optimize the performance.

\section{Extensions}
\label{sec:discu}

\subsection{Beyond QAOA}

While the primary focus of this paper is on optimizing QAOA on bipotent architectures without requiring additional calibration, the insights presented here have broader implications for other quantum algorithms and architectures. Specifically, some two-qubit gates in an algorithm may only achieve high fidelity after a time-consuming gate calibration, resulting in a notable increase in overhead costs \cite{lao2021designing}. However, our study suggests a \emph{selective calibration} approach that can reduce the potential overhead by calibrating only the two-qubit gates implemented on specific qubit pairs with a higher CX gate error rate, instead of calibrating all two-qubit gates. This approach enhances algorithm performance by enabling more efficient resource utilization, leading to faster and more accurate results. Moreover, pulse-level optimization and various gate decomposition strategies can be combined to significantly improve algorithm performance. By investigating the resilience of different pulse shapes to noise, we can identify more robust pulse shapes that are less affected by noise and errors in the computation. Consequently, by leveraging these resilient pulse shapes, we can achieve better performance of quantum algorithms.

We optimize the algorithm on cross-resonance based hardware. The optimization is also applicable to other architectures, such as tunable couplers \cite{McKay2016} that support interactions like iSWAP and CZ gates \cite{Ganzhorn2020, Sung2021}. The implementation of ZZ or ZZ-SWAP gates can be realized using three sets of gates: CZ gates, iSWAP gates, and a combination of CZ and iSWAP gates. By tailoring distinct gate decompositions to the properties of the two-qubit gates on the target hardware, such as fidelity and schedule duration, we can achieve improved performance. Furthermore, integrating selective calibration at the pulse-level makes this approach particularly promising for near-term quantum computing. Additionally, our study shows that decomposing ZZ-SWAP gates into CZ gates helps achieve improved results. Therefore, we expect that architectures with a native CZ gate may benefit even more from this optimization. Importantly, these techniques are not restricted to a particular algorithm and can be employed in various other quantum algorithms.

\subsection{Error Mitigation}

In this paper, we applied readout error mitigation in Qiskit to mitigate measurement errors. However, there are additional techniques that can be utilized to suppress other types of errors in bipotent architectures. Here, we discuss two such techniques: dynamical decoupling \cite{Pokharel2018} and zero-noise extrapolation (ZNE) \cite{Temme2017}.

Dynamical decoupling is a widely-used strategy to suppress decoherence errors \cite{Viola1998} by applying pulse sequences to idle qubits. In bipotent architectures, the presence of two distinct CX gate types with unique pulse schedules can lead to significant variations in pulse lengths across different qubit pairs within the algorithm. Applying dynamical decoupling pulses to the idle regions of shorter pulses can be particularly advantageous for such architectures. Moreover, diverse gate decompositions on different qubit pairs and selective calibration can also result in noticeable differences in pulse lengths and larger idle regions. This underscores the importance of incorporating dynamical decoupling into this optimization strategy.

The ZNE technique estimates the noiseless result by utilizing expectation values measured at different noise levels. In bipotent architectures, two types of CX gates introduce different errors, which can be suppressed using ZNE. Additionally, ZNE can help identify the optimal gate decomposition for each qubit pair among various decomposition strategies used on different qubit pairs.

Thorough analysis, experimentation, and adaptation of these techniques to the unique characteristics of bipotent architectures are crucial for achieving improved performance on such architectures.

\section{Conclusions and Future Work}
\label{sec:concl}

Bipotent QPU architectures, which offer improved functions on some but not all of their qubits, provide new opportunities for quantum computations, but they also introduce new complexities. In this paper, we have investigated the choice between improved direct-CX gates with better fidelity and duration and conventional ECR-CX gates that support pulse-level optimizations without additional calibration. Based on a careful validation of different circuits on two IBM's QPUs, we found that ECR-CX with pulse-level optimizations used to a maximum extent outperforms even seemingly globally optimal circuits with the highest fidelity. When comparing two approaches to constructing bipotent circuits, Bipotent-circuits that consider both fidelity and duration dominate Global-circuits that focus only on fidelity.

We believe that the hardware-software codesign is an essential issue in emerging bipotent architectures. For instance, the QAOA circuits investigated in this paper have a regular structure with relatively few quantum gate types and well-understood pulse-level optimizations. For other types of circuits with less pronounced effects of pulse-level optimizations, the improvements due to better basis gates may be more advantageous. An interesting avenue for future work is the development of universal and fully automatic methods to map a given quantum circuit to a bipotent QPU, thereby maximizing its probability of success and minimizing the duration of the resulting schedule. We believe that the existence of such methods will be a prerequisite for the practical success of bipotent quantum architectures. It would also be interesting to explore the potential applicability of the proposed method on other QPUs such as Google's Sycamore processor \cite{arute2019quantum} with a two-qubit hardware-native gate set \{CZ, $\mathrm{\sqrt{iSWAP}}$, Sycamore\}.

Combining the proposed optimization strategies with error characterization, selective calibration, error correction, and various error mitigation techniques would be valuable in further enhancing the performance of algorithms on bipotent quantum architectures. Error characterization allows for a comprehensive understanding of the specific noise sources and patterns within the system. Selective calibration fine-tunes individual qubit pairs based on their unique characteristics, optimizing their performance. Error correction protects quantum information from errors caused by interactions with the environment \cite{Bennett1996}. Additionally, error mitigation suppresses errors and improves performance. We have discussed the potential benefits of utilizing dynamical decoupling and ZNE based on the properties of the two CX gate types specifically for bipotent architectures. Integrating these techniques improves the performance and reliability of algorithms on bipotent quantum architectures, enabling more accurate and efficient quantum computations.

\begin{acknowledgments}
The authors would like to thank Joris Kattemölle and Xi Chen for their constructive criticism of the manuscript. The authors would also like to thank Thomas Wellens, John P. Hayes, Sebastian Brandhofer, and Andreas Ketterer for their useful discussions. This work was supported in part by the  project QORA within the Competence Center Quantum Computing Baden-W\"urttemberg (funded by the  Ministerium f\"ur Wirtschaft, Arbeit und Wohnungbau Baden-W\"urttemberg).
\end{acknowledgments}

\appendix*

\section{Cloud Platform Details}

Here we provide more details about the two cloud-accessible QPUs ibmq\_ehningen and ibm\_auckland used in our research. Both QPUs have identical layouts, as shown in Figs.~\ref{fig:gate_map_ehningen} and \ref{fig:gate_map_auckland}. The basis gates of both QPUs include ID, RZ, SX, X, and CX, where ID is the identity gate, X is the Pauli X gate, and SX is the square root of the X gate. Calibration data at the time of the demonstration of QAOA for PortOpt and MaxCut are summarized in Tables \ref{table:calibration_data_portopt} and \ref{table:calibration_data_maxcut}, respectively.

\begin{table*}
\begin{ruledtabular}
  \centering
  \caption{Calibration data at the time of the demonstration presented in Sec. \ref{subsubsec:bench_qaoa_portopt}.}
  \label{table:calibration_data_portopt}
  \begin{tabular}{lllllll}
   \textbf{3Q-QAOA-PortOpt}&\multicolumn{2}{l}{\textbf{ECR-circuit}}&\multicolumn{2}{l}{\textbf{Global-D}}&\multicolumn{2}{l}{\textbf{Bipotent-circuit}}\\
    Qubits used & \multicolumn{2}{l}{[16, 14, 11]} & \multicolumn{2}{l}{[18, 21, 23]} & \multicolumn{2}{l}{[22, 25, 26]}\\
    $T_1~(\rm{\mu s})$ & \multicolumn{2}{l}{[194.12, 163.4, 150.43]} & \multicolumn{2}{l}{[214.69, 300.02, 222.14]} & \multicolumn{2}{l}{[195.66, 257.93, 195.59]}\\
    $T_2~(\rm{\mu s})$ & \multicolumn{2}{l}{[240.94, 215.18, 174.7]} & \multicolumn{2}{l}{[333.28, 155.0, 234.21]} & \multicolumn{2}{l}{[31.47, 354.96, 27.85]}\\
    Frequency (GHz) & \multicolumn{2}{l}{[5.022, 5.177, 5.119]} & \multicolumn{2}{l}{[4.996, 4.94, 4.805]} & \multicolumn{2}{l}{[4.725, 4.95, 5.151]}\\
    Anharmonicity (GHz) & \multicolumn{2}{l}{[-0.3435, -0.3408, -0.3405]} & \multicolumn{2}{l}{[-0.343, -0.3456, -0.3471]} & \multicolumn{2}{l}{[-0.3464, -0.3457, -0.3391]}\\
    Prob. meas. 0 prep. $\ket{1}$ (\%) & \multicolumn{2}{l}{[0.64, 0.92, 1.66]} & \multicolumn{2}{l}{[1.14, 1.0, 0.84]} & \multicolumn{2}{l}{[1.4, 1.16, 0.8]}\\
    Prob. meas. 1 prep. $\ket{0}$ (\%) & \multicolumn{2}{l}{[0.56, 0.68, 1.4]} & \multicolumn{2}{l}{[0.86, 0.78, 0.76]} & \multicolumn{2}{l}{[1.12, 0.66, 0.66]}\\
    Readout length (ns) & \multicolumn{2}{l}{[846.22, 846.22, 846.22]} & \multicolumn{2}{l}{[846.22, 846.22, 846.22]} & \multicolumn{2}{l}{[846.22, 846.22, 846.22]}\\
    Readout error (\%) & \multicolumn{2}{l}{[0.6, 0.8, 1.53]} & \multicolumn{2}{l}{[1.0, 0.89, 0.8]} & \multicolumn{2}{l}{[1.26, 0.91, 0.73]}\\
    Single-qubit gate error (\%) & \multicolumn{2}{l}{[0.016, 0.033, 0.022]} & \multicolumn{2}{l}{[0.026, 0.021, 0.013]} & \multicolumn{2}{l}{[0.016, 0.024, 0.014]}\\
    CX gate error (\%) & \multicolumn{2}{l}{[0.66, 0.89]} & \multicolumn{2}{l}{[0.44, 0.52]} & \multicolumn{2}{l}{[0.62, 0.57]}\\
    \\
    \textbf{4Q-QAOA-PortOpt}&\multicolumn{2}{l}{\textbf{ECR-circuit}} & \multicolumn{2}{l}{\textbf{Global-B}} & \multicolumn{2}{l}{\textbf{Direct-circuit}}\\
    Qubits used & \multicolumn{2}{l}{[16, 14, 11, 8]} & \multicolumn{2}{l}{[18, 21, 23, 24]} & \multicolumn{2}{l}{[1, 4, 7, 6]}\\
    $T_1~(\rm{\mu s})$ & \multicolumn{2}{l}{[194.12, 163.4, 150.43, 145.9]} & \multicolumn{2}{l}{[214.69, 300.02, 222.14, 244.38]} & \multicolumn{2}{l}{[243.57, 134.77, 205.84, 105.37]}\\
    $T_2~(\rm{\mu s})$ & \multicolumn{2}{l}{[240.94, 215.18, 174.7, 157.35]} & \multicolumn{2}{l}{[333.28, 155.0, 234.21, 225.4]} & \multicolumn{2}{l}{[172.38, 111.81, 266.71, 183.16]}\\
    Frequency (GHz) & \multicolumn{2}{l}{[5.022, 5.177, 5.119, 5.174]} & \multicolumn{2}{l}{[4.996, 4.94, 4.805, 5.074]} & \multicolumn{2}{l}{[5.182, 5.054, 4.978, 4.89]}\\
    Anharmonicity (GHz) & \multicolumn{2}{l}{[-0.3435, -0.3408, -0.3405, -0.3399]} & \multicolumn{2}{l}{[-0.343, -0.3456, -0.3471, -0.3416]} & \multicolumn{2}{l}{[-0.34, -0.3426, -0.344, -0.3448]}\\
    Prob. meas. 0 prep. $\ket{1}$ (\%) & \multicolumn{2}{l}{[0.64, 0.92, 1.66, 1.18]} & \multicolumn{2}{l}{[1.14, 1.0, 0.84, 0.88]} & \multicolumn{2}{l}{[1.04, 1.02, 0.74, 1.48]}\\
    Prob. meas. 1 prep. $\ket{0}$ (\%) & \multicolumn{2}{l}{[0.56, 0.68, 1.4, 1.46]} & \multicolumn{2}{l}{[0.86, 0.78, 0.76, 0.66]} & \multicolumn{2}{l}{[0.64, 0.58, 0.82, 2.02]}\\
    Readout length (ns) & \multicolumn{2}{l}{[846.22, 846.22, 846.22, 846.22]} & \multicolumn{2}{l}{[846.22, 846.22, 846.22, 846.22]} & \multicolumn{2}{l}{[846.22, 846.22, 846.22, 846.22]}\\
    Readout error (\%) & \multicolumn{2}{l}{[0.6, 0.8, 1.53, 1.32]} & \multicolumn{2}{l}{[1.0, 0.89, 0.8, 0.77]} & \multicolumn{2}{l}{[0.84, 0.8, 0.78, 1.75]}\\
    Single-qubit gate error (\%) & \multicolumn{2}{l}{[0.016, 0.033, 0.022, 0.029]} & \multicolumn{2}{l}{[0.026, 0.021, 0.013, 0.017]} & \multicolumn{2}{l}{[0.026, 0.025, 0.016, 0.02]}\\
    CX gate error (\%) & \multicolumn{2}{l}{[0.66, 0.89, 1.06]} & \multicolumn{2}{l}{[0.44, 0.52, 0.49]} & \multicolumn{2}{l}{[0.68, 0.4, 0.35]}\\
    \\
    \textbf{5Q-QAOA-PortOpt}&\multicolumn{3}{l}{\textbf{ECR-circuit}} & \multicolumn{3}{l}{\textbf{Global-B}}\\
    Qubits used & \multicolumn{3}{l}{[16, 14, 11, 8, 9]} & \multicolumn{3}{l}{[17, 18, 21, 23, 24]}\\
    $T_1~(\rm{\mu s})$ & \multicolumn{3}{l}{[194.12, 163.4, 150.43, 145.9, 178.39]} & \multicolumn{3}{l}{[143.0, 214.69, 300.02, 222.14, 244.38]}\\
    $T_2~(\rm{\mu s})$ & \multicolumn{3}{l}{[240.94, 215.18, 174.7, 157.35, 195.4]} & \multicolumn{3}{l}{[39.76, 333.28, 155.0, 234.21, 225.4]}\\
    Frequency (GHz) & \multicolumn{3}{l}{[5.022, 5.177, 5.119, 5.174, 4.993]} & \multicolumn{3}{l}{[5.136, 4.996, 4.94, 4.805, 5.074]}\\
    Anharmonicity (GHz) & \multicolumn{3}{l}{[-0.3435, -0.3408, -0.3405, -0.3399, -0.3441]} & \multicolumn{3}{l}{[-0.341, -0.343, -0.3456, -0.3471, -0.3416]}\\
    Prob. meas. 0 prep. $\ket{1}$ (\%) & \multicolumn{3}{l}{[0.64, 0.92, 1.66, 1.18, 1.0]} & \multicolumn{3}{l}{[0.88, 1.14, 1.0, 0.84, 0.88]}\\
    Prob. meas. 1 prep. $\ket{0}$ (\%) & \multicolumn{3}{l}{[0.56, 0.68, 1.4, 1.46, 0.74]} & \multicolumn{3}{l}{[0.56, 0.86, 0.78, 0.76, 0.66]}\\
    Readout length (ns) & \multicolumn{3}{l}{[846.22, 846.22, 846.22, 846.22, 846.22]} & \multicolumn{3}{l}{[846.22, 846.22, 846.22, 846.22, 846.22]}\\
    Readout error (\%) & \multicolumn{3}{l}{[0.6, 0.8, 1.53, 1.32, 0.87]} & \multicolumn{3}{l}{[0.72, 1.0, 0.89, 0.8, 0.77]}\\
    Single-qubit gate error (\%) & \multicolumn{3}{l}{[0.016, 0.033, 0.022, 0.029, 0.02]} & \multicolumn{3}{l}{[0.018, 0.026, 0.021, 0.013, 0.017]}\\
    CX gate error (\%) & \multicolumn{3}{l}{[0.66, 0.89, 1.06, 0.94]} & \multicolumn{3}{l}{[0.47, 0.44, 0.52, 0.49]}\\
    \\
    \textbf{5Q-QAOA-PortOpt} & \multicolumn{3}{l}{\textbf{Direct-circuit}} & \multicolumn{3}{l}{\textbf{Bipotent-circuit}}\\
    Qubits used & \multicolumn{3}{l}{[6, 7, 4, 1, 2]} & \multicolumn{3}{l}{[11, 14, 13, 12, 10]}\\
    $T_1~(\rm{\mu s})$ & \multicolumn{3}{l}{[42.89, 148.29, 108.74, 154.56, 75.75]} & \multicolumn{3}{l}{[127.56, 107.65, 147.9, 183.56, 115.74]}\\
    $T_2~(\rm{\mu s})$ & \multicolumn{3}{l}{[183.16, 266.71, 111.81, 172.38, 19.75]} & \multicolumn{3}{l}{[74.74, 196.1, 271.77, 442.33, 57.65]}\\
    Frequency (GHz) & \multicolumn{3}{l}{[4.89, 4.978, 5.054, 5.182, 5.127]} & \multicolumn{3}{l}{[5.119, 5.177, 4.926, 4.725, 4.835]}\\
    Anharmonicity (GHz) & \multicolumn{3}{l}{[-0.3448, -0.344, -0.3426, -0.34, -0.3403]} & \multicolumn{3}{l}{[-0.3405, -0.3408, -0.344, -0.3484, -0.3471]}\\
    Prob. meas. 0 prep. $\ket{1}$ (\%) & \multicolumn{3}{l}{[1.48, 0.74, 1.02, 1.04, 1.04]} & \multicolumn{3}{l}{[1.54, 1.08, 1.74, 1.0, 0.74]}\\
    Prob. meas. 1 prep. $\ket{0}$ (\%) & \multicolumn{3}{l}{[2.02, 0.82, 0.58, 0.64, 0.66]} & \multicolumn{3}{l}{[1.34, 0.44, 0.96, 0.9, 0.46]}\\
    Readout length (ns) & \multicolumn{3}{l}{[846.22, 846.22, 846.22, 846.22, 846.22]} & \multicolumn{3}{l}{[846.22, 846.22, 846.22, 846.22, 846.22]}\\
    Readout error (\%) & \multicolumn{3}{l}{[1.75, 0.78, 0.8, 0.84, 0.85]} & \multicolumn{3}{l}{[1.44, 0.76, 1.35, 0.95, 0.6]}\\
    Single-qubit gate error (\%) & \multicolumn{3}{l}{[0.02, 0.016, 0.025, 0.026, 0.05]} & \multicolumn{3}{l}{[0.025, 0.031, 0.018, 0.022, 0.017]}\\
    CX gate error (\%) & \multicolumn{3}{l}{[0.35, 0.4, 0.68, 0.96]} & \multicolumn{3}{l}{[0.84, 0.59, 0.66, 0.5]}\\
    \\
    \textbf{5Q-QAOA-PortOpt} & \multicolumn{3}{l}{\textbf{Bipotent-circuit (ibm\_auckland)}}\\
    Qubits used & \multicolumn{3}{l}{[11, 8, 5, 3, 2]}\\
    $T_1~(\rm{\mu s})$ & \multicolumn{3}{l}{[168.1, 177.64, 280.77, 157.44, 126.14]}\\
    $T_2~(\rm{\mu s})$ & \multicolumn{3}{l}{[138.88, 97.6, 96.99, 219.48, 196.01]}\\
    Frequency (GHz) & \multicolumn{3}{l}{[5.055, 5.204, 4.993, 4.897, 5.006]}\\
    Anharmonicity (GHz) & \multicolumn{3}{l}{[-0.3422, -0.3407, -0.3445, -0.3455, -0.3434]}\\
    Prob. meas. 0 prep. $\ket{1}$ (\%) & \multicolumn{3}{l}{[0.84, 1.08, 1.14, 1.48, 1.28]}\\
    Prob. meas. 1 prep. $\ket{0}$ (\%) & \multicolumn{3}{l}{[0.5, 0.7, 0.74, 0.84, 0.98]}\\
    Readout length (ns) & \multicolumn{3}{l}{[757.33, 757.33, 757.33, 757.33, 757.33]}\\
    Readout error (\%) & \multicolumn{3}{l}{[0.67, 0.89, 0.94, 1.16, 1.13]}\\
    Single-qubit gate error (\%) & \multicolumn{3}{l}{[0.017, 0.019, 0.049, 0.016, 0.018]}\\
    CX gate error (\%) & \multicolumn{3}{l}{[0.63, 0.98, 0.59, 0.49]}
  \end{tabular}
\end{ruledtabular}
\end{table*}

\begin{table*}
\begin{ruledtabular}
  \centering
  \caption{Calibration data at the time of the demonstration presented in Sec. \ref{subsubsec:bench_qaoa_maxcut}.}
  \label{table:calibration_data_maxcut}
  \begin{tabular}{lll}
    
    \textbf{5Q-QAOA-MaxCut}&\textbf{ECR-circuit}&\textbf{Global-B}\\
    Qubits used & [16, 14, 11, 8, 9] & [6, 7, 4, 1, 0]\\
    $T_1~(\rm{\mu s})$ & [197.21, 207.15, 104.89, 122.12, 135.3] & [186.94, 217.29, 129.35, 143.59, 197.13]\\
    $T_2~(\rm{\mu s})$ & [240.94, 215.18, 174.7, 157.35, 195.4] & [183.16, 266.71, 111.81, 172.38, 201.31]\\
    Frequency (GHz) & [5.022, 5.177, 5.119, 5.174, 4.993] & [4.89, 4.978, 5.054, 5.182, 4.961] \\
    Anharmonicity (GHz) & [-0.3435, -0.3408, -0.3405, -0.3399, -0.3441] & [-0.3448, -0.344, -0.3426, -0.34, -0.3445]\\
    Prob. meas. 0 prep. $\ket{1}$ (\%) & [0.64, 0.92, 1.66, 1.18, 1.0] & [1.48, 0.74, 1.02, 1.04, 1.04]\\
    Prob. meas. 1 prep. $\ket{0}$ (\%) & [0.56, 0.68, 1.4, 1.46, 0.74] & [2.02, 0.82, 0.58, 0.64, 0.88]\\
    Readout length (ns) & [846.22, 846.22, 846.22, 846.22, 846.22] & [846.22, 846.22, 846.22, 846.22, 846.22]\\
    Readout error (\%) & [0.6, 0.8, 1.53, 1.32, 0.87] & [1.75, 0.78, 0.8, 0.84, 0.96] \\
    Single-qubit gate error (\%) & [0.016, 0.033, 0.022, 0.029, 0.02] & [0.02, 0.016, 0.025, 0.026, 0.018]\\
    CX gate error (\%) & [0.66, 0.89, 1.06, 0.94] & [0.35, 0.4, 0.68, 0.51]\\
    \\
    \textbf{5Q-QAOA-MaxCut} &\textbf{Direct-circuit}&\textbf{Bipotent-circuit}\\
    Qubits used  & [6, 7, 4, 1, 2] & [17, 18, 21, 23, 24]\\
    $T_1~(\rm{\mu s})$ & [42.89, 148.29, 108.74, 154.56, 75.75] & [143.0, 214.69, 300.02, 222.14, 244.38]\\
    $T_2~(\rm{\mu s})$ & [183.16, 266.71, 111.81, 172.38, 19.75] & [39.76, 333.28, 155.0, 234.21, 225.4]\\
    Frequency (GHz) & [4.89, 4.978, 5.054, 5.182, 5.127] & [5.136, 4.996, 4.94, 4.805, 5.074]\\
    Anharmonicity (GHz) & [-0.3448, -0.344, -0.3426, -0.34, -0.3403] & [-0.341, -0.343, -0.3456, -0.3471, -0.3416]\\
    Prob. meas. 0 prep. $\ket{1}$ (\%) & [1.48, 0.74, 1.02, 1.04, 1.04] & [0.88, 1.14, 1.0, 0.84, 0.88]\\
    Prob. meas. 1 prep. $\ket{0}$ (\%) & [2.02, 0.82, 0.58, 0.64, 0.66] & [0.56, 0.86, 0.78, 0.76, 0.66]\\
    Readout length (ns) & [846.22, 846.22, 846.22, 846.22, 846.22] & [846.22, 846.22, 846.22, 846.22, 846.22]\\
    Readout error (\%) & [1.75, 0.78, 0.8, 0.84, 0.85] & [0.72, 1.0, 0.89, 0.8, 0.77]\\
    Single-qubit gate error (\%) & [0.02, 0.016, 0.025, 0.026, 0.05] & [0.018, 0.026, 0.021, 0.013, 0.017]\\
    CX gate error (\%) & [0.35, 0.4, 0.68, 0.96] & [0.47, 0.44, 0.52, 0.49]
  \end{tabular}
\end{ruledtabular}
\end{table*}

\nocite{*}

\bibliography{refs}

\end{document}